\documentstyle[12pt,leqno]{article}
\def\DATE{November 28, 1994}

\newtheorem{theorem}{Theorem}[section]
\newtheorem{definition}[theorem]{Definition}
\newtheorem{observation}[theorem]{Observation}
\newtheorem{odstavec}[theorem]{}
\newtheorem{example}[theorem]{Example}
\newtheorem{lemma}[theorem]{Lemma}
\newtheorem{proposition}[theorem]{Proposition}

\catcode`\@=11
\def\@begintheorem#1#2{\it \trivlist \item[\hskip
 \labelsep{\bf #1\ #2.}]}
\def\@opargbegintheorem#1#2#3{\it \trivlist\item[\hskip%
 \labelsep{\bf #1\ #2.\ (#3)}]}
\def\@endtheorem{\endtrivlist}

\def\@listI{\leftmargin\leftmargini \parsep 1pt plus 2.5pt
 minus 1pt\topsep 5pt plus 2pt minus 3pt%
 \itemsep 0pt plus 2.5pt minus 1pt}
\let\@listi\@listI
\@listi

\def\@sect#1#2#3#4#5#6[#7]#8{\ifnum #2>\c@secnumdepth%
 \def \@svsec {}\else \refstepcounter {#1}\edef \@svsec%
 {\csname the#1\endcsname. \hskip .1em }\fi \@tempskipa%
 #5\relax \ifdim \@tempskipa >\z@ \begingroup #6\relax%
 \@hangfrom {\hskip #3\relax \@svsec }{\interlinepenalty%
 \@M #8.\par }\endgroup \csname #1mark\endcsname {#7}%
 \addcontentsline {toc}{#1}{\ifnum #2>\c@secnumdepth%
 \else \protect \numberline {\csname the#1\endcsname. }%
 \fi #7}\else \def \@svsechd {#6\hskip #3\@svsec #8.%
 \csname #1mark\endcsname {#7}\addcontentsline {toc}{#1}%
 {\ifnum #2>\c@secnumdepth \else \protect \numberline%
 {\csname the#1\endcsname. }\fi #7}}\fi \@xsect {#5}}

\def\section{\@startsection {section}{1}{\z@ }%
 {-3.5ex plus -1ex minus -.2ex}{2.3ex plus .2ex}{\bf }}

\def\abstract{%
\if@twocolumn \section *{Abstract}
 \else \small\quotation\noindent{\bf Abstract.}\fi}

\def\thebibliography#1{%
 \section *{References.\@mkboth {REFERENCES}{REFERENCES}}%
 \list {[\arabic {enumi}]}{\settowidth \labelwidth {[#1]}%
 \leftmargin \labelwidth \advance \leftmargin \labelsep %
 \usecounter {enumi}} \def \newblock %
 {\hskip .11em plus .33em minus -.07em} \sloppy \clubpenalty 4000%
 \widowpenalty 4000 \sfcode`\.=1000\relax}

\def\@maketitle{%
 \newpage \null \vskip 2em
 \begin{center}{\Large\bf \@title \par }
 \vskip 1.5em
 {\large \lineskip .5em
 \begin {tabular}[t]{c}\@author
 \end{tabular}\par } \vskip .8em {\DATE}
 \end{center}\par \vskip 1.5em}

\catcode`\@=13

\newcommand{\Ytriangle}[6]{
\setlength{\unitlength}{.55cm}
\begin{picture}(6,4.1)(0,-.1)
\thicklines

\put(0,3.5){\makebox(0,0){$#1$}}
\put(3.5,1.75){\makebox(0,0){$#2$}}
\put(0,0){\makebox(0,0){$#3$}}

\put(0,3){\vector(0,-1){2.5}}
\put(1,0.5){\vector(2,1){2}}
\put(1,3){\vector(2,-1){2}}

\put(-.5,1.75){\makebox(0,0)[r]{$#4$}}
\put(1.75,.5){\makebox(0,0)[l]{$#5$}}
\put(1.75,3){\makebox(0,0)[l]{$#6$}}
\end{picture}
}

\newcommand{\Ztriangle}[6]{
\setlength{\unitlength}{.7cm}
\begin{picture}(6,4.1)(0,-.1)
\thicklines

\put(0,3.5){\makebox(0,0){$#1$}}
\put(4.5,1.75){\makebox(0,0){$#2$}}
\put(0,0){\makebox(0,0){$#3$}}

\put(0,3){\vector(0,-1){2.5}}
\put(1,0.5){\vector(2,1){2}}
\put(1,3){\vector(2,-1){2}}

\put(-.5,1.75){\makebox(0,0)[r]{$#4$}}
\put(1.75,.5){\makebox(0,0)[l]{$#5$}}
\put(1.75,3){\makebox(0,0)[l]{$#6$}}
\end{picture}
}

\newcommand{\Htriangle}[6]{
\setlength{\unitlength}{.55cm}
\begin{picture}(6,4.1)(0,-.1)
\thicklines
\put(0,3){\makebox(0,0){$#1$}}
\put(6,3){\makebox(0,0){$#2$}}
\put(6,0){\makebox(0,0){$#3$}}

\put(1,3){\vector(1,0){3.5}}
\put(6,2.5){\vector(0,-1){2}}
\put(1,2.5){\vector(2,-1){4}}

\put(6.3,1.5){\makebox(0,0)[l]{$#5$}}
\put(3,3.5){\makebox(0,0)[b]{$#4$}}
\put(2.8,1.25){\makebox(0,0)[r]{$#6$}}
\end{picture}
}

\newcommand{\square}[8]{
\setlength{\unitlength}{.55cm}
\begin{picture}(5,3.6)
\thicklines

\put(0,3){\makebox(0,0){$#1$}}
\put(5,3){\makebox(0,0){$#2$}}
\put(0,0){\makebox(0,0){$#3$}}
\put(5,0){\makebox(0,0){$#4$}}

\put(-.5,1.5){\makebox(0,0)[r]{$#6$}}
\put(5.5,1.5){\makebox(0,0)[l]{$#7$}}
\put(2.5,0.5){\makebox(0,0)[b]{$#8$}}
\put(2.5,3.5){\makebox(0,0)[b]{$#5$}}

\put(1.3,0){\vector(1,0){2.7}}
\put(1,3){\vector(1,0){3}}
\put(0,2.5){\vector(0,-1){2}}
\put(5,2.5){\vector(0,-1){2}}
\end{picture}
}

\newcommand{\Square}[8]{
\setlength{\unitlength}{.55cm}
\begin{picture}(5,3.6)
\thicklines

\put(0,3){\makebox(0,0){$#1$}}
\put(5,3){\makebox(0,0){$#2$}}
\put(0,0){\makebox(0,0){$#3$}}
\put(5,0){\makebox(0,0){$#4$}}

\put(-.5,1.5){\makebox(0,0)[r]{$#6$}}
\put(5.5,1.5){\makebox(0,0)[l]{$#7$}}
\put(2.5,0.5){\makebox(0,0)[b]{$#8$}}
\put(2.5,3.5){\makebox(0,0)[b]{$#5$}}

\put(1.6,0){\vector(1,0){2.3}}
\put(1,3){\vector(1,0){3}}
\put(0,2.5){\vector(0,-1){2}}
\put(5,2.5){\vector(0,-1){2}}
\end{picture}
}

\def\qed{\hspace*{\fill}\mbox{\hphantom{mm}\rule{0.25cm}{0.25cm}}\\}
\def\calp{{\cal P}} \def\cali{{\cal I}} \def\cala{{\cal A}}
\def\calk{{\cal K}} \def\calh{{\cal H}} \def\calc{{\cal C}}
\def\calt{{\cal T}} \def\calr{{\cal R}}
\def\calo{{\cal O}} \def\cale{{\cal E}}
\def\caly{{\cal Y}} \def\calu{{\cal U}}
\def\caln{{\cal N}} \def\calm{{\cal M}}
\def\calf{{\cal F}} \def\cald{{\cal D}}
\def\cals{{\cal S}} \def\S{$\cals$}
\def\barcalp{{\overline{\calp}}}

\def\hatcals{{\hat{\cals}}}
\def\ot{\otimes}
\def\bk{{\bf k}}
\def\comp{\circ}
\def\Im{\mbox{\rm Im}}
\def\Ker{\mbox{\rm Ker}}
\def\id{1\!\!1}
\def\free{{\cal F}}
\def\oper{{\bf Oper}}
\def\coll{{\bf Coll}}
\def\der{{\mbox{\rm Der}}}
\def\Der{{\mbox{\rm Der}}}
\def\smod{\mbox{$\cals$-{\bf Mod}}}
\def\frees#1{{\cals\langle {#1}\rangle}}
\def\End{\mbox{\it End\/}}
\def\op{\oplus}
\def\susp{\uparrow\!}
\def\desusp{\downarrow\!}

\def\freez#1{\free(Z^{{#1}})}
\def\cl{\mbox{\rm cl}}
\def\overto#1{\stackrel{#1}{\longrightarrow}}
\def\proj{p}
\def\Span{{\mbox{\rm Span}}}
\def\csm#1#2{C^{#1,#2}(\cals;M)}
\def\ssusp{{\bf s}}
\def\dend{{d_{End}}}
\def\cot#1#2{{T^{#1,#2}}}
\def\cotsm#1#2{{\cot{#1}{#2}(\cals;M)}}
\def\comm{{\it Comm}}
\def\ass{{\it Ass}}
\def\st{{\widetilde\cals}}
\def\fint{{\free_{\bf Z}}}
\def\uN{\underline{\cal N}}
\def\uM{\underline{\cal M}}
\def\cN{{\cal C}_N}

\begin{document}
\baselineskip18pt

\title{Models for operads}
\author{Martin Markl\thanks{Partially
supported by the National Research Counsel, USA}}

\maketitle

\begin{abstract}
We study properties of differential graded (dg) operads modulo weak
equivalences, that is, modulo the relation given by the existence of a
chain of dg operad maps inducing a homology isomorphism. This
approach, naturally arising in string theory, leads us to consider
various versions of models. Some applications in topology
(homotopy-everything spaces), algebra (cotangent cohomology) and
mathematical physics (closed string-field theory) -- are also given.

\vskip3pt
\noindent
{{\bf Classification:} 18C10, 55P62}

\end{abstract}

\noindent{\bf Introduction.}
Our aim is to show that some basic constructions of
rational homotopy theory (such as minimal models, bigraded and
filtered models, etc.) are available also for the category
of operads and to indicate also some
applications of this fact in topology, homological algebra and
mathematical physics.
At the very end of the paper we also try to convince the
reader that the ``homotopy theory of
operads'' naturally arises in some situations of string theory. We
also hope that the paper will help to understand better the (co)bar
construction for operads and the related property of Koszulness
introduced in~\cite{GK}.

Our constructions are related with a
closed model category structure on the category of operads whose
existence is more or less known.
A stunning observation is that our constructions
are even easier than the corresponding constructions in rational
homotopy theory. Our explanation is that, while in rational
homotopy theory we work basically with graded objects (algebras of
various types), the nature of operads resembles more {\em bigraded\/}
algebras, one grading (inner) being given by the grading of the
underlying vector space, the other given by the ``number of
inputs''. Experience then says ``the more independent gradings, the
better'', which would certainly resonate for anyone familiar with
the construction of a trigraded model of~\cite{F}.

The concept presented here was originated in~\cite{M} where we tried
to do ``homological algebra for {\small PROP}s'' (operad is an
algebraic {\small PROP}). It was then the
illuminating paper~\cite{GK} which made us believe that the
constructions presented here were possible.

As far as our constructions are concerned,
we almost completely neglect the
uniqueness problem, which would require a notion of homotopy for the
category of operads and we are afraid that this would stretch the
length of the paper beyond any reasonable limit. Fortunately, our
applications are mostly independent on this assumption.

All our constructions are primarily made for (unital) operads $\cals$
with
$\cals(1)=$ the ground field, which is always supposed to be of
characteristic zero. This assumption plays here the
r\^ole of simple connectivity in rational homotopy theory. We will
see that for any (unital) augmented operad $\cals$ there exists an
operad
$\widetilde\cals$ satisfying this assumption,
constructed from $\cals$ in a canonical way, which may be
understood as an analog of the universal covering of a topological
space, and we may apply our constructions to $\widetilde\cals$ instead
of $\cals$. Even this could give interesting results.

Central for the {\em applications\/} of our theory is
Proposition~\ref{stunning} which gives an easy criterion for
intrinsic formality of an operad. Note that there is no analog of
this result in rational homotopy theory. From this statement we
deduce, for example, that a closed string-field theory induces a
homotopy Lie algebra structure on the space of relative states
(Example~\ref{doutnicek}, Proposition~\ref{prstynek}), which
is one of main results of~\cite{KSV}. For more applications, see
par.~5.

Another by-product of our theory is a new, very general definition of
homotopy versions of algebraic objects -- homotopy $\cals$-algebra is
an algebra over the minimal model of $\cals$, see par.~5.
For Koszul operads our definition coincides with the
definition of~\cite{GK} (Example~\ref{ferda}), but it is not
restricted to the Koszul case, see Example~\ref{fuk}.

The paper is not the only example of possible applications
of the ``rational way of thinking'' in other branches of mathematics;
let us mention at least the applications in local algebra,
see~\cite{AH}.

We would like to express our thanks to J.D.~Stasheff for careful
reading of the manuscript and for many valuable suggestions and
remarks.

Plan of the paper:
\begin{center}
\begin{minipage}{15cm}
\baselineskip18pt
\begin{enumerate}
\item
Preliminaries.
\item
Self-dual nature of the category of operads.
\item
Models for operads.
\item
Cotangent cohomology.
\item
Homotopy $\cals$-algebras, homotopy everything spaces.
\end{enumerate}
\end{minipage}
\end{center}

\section{Preliminaries}

All algebraic objects in the paper will be considered over a fixed
field $\bk$ of characteristic zero.
For a graded vector space $V=\bigoplus_p V_p$ let $\susp V$ (resp.
$\desusp V$) be the suspension (resp. the desuspension) of $V$, i.e.
the graded
vector space defined by $(\susp V)_p = V_{p-1}$ (resp. $(\desusp V)_p
= V_{p+1}$).
We have the obvious natural maps $\uparrow : V \to \susp V$ and
$\downarrow: V\to \desusp V$.
By $\#V$ we denote the dual of $V$, i.e. the graded
vector space with $(\#V)_p$ := the space of linear maps $\phi :V_p
\to \bk$.
If, moreover, an action of the
symmetric group $\Sigma_n$ on $V$ is given, we will consider the dual
$\#V$
with the action given by $(\sigma \phi)(v) :=
(-1)^{\mbox{\scriptsize sgn}(\sigma)}\phi(\sigma^{-1}v)$, for $\sigma
\in \Sigma$,
$\phi \in \#V$ and $v\in V$.

Let $\calp$ be a (graded) operad in the usual sense
(=~an operad in the monoidal category of graded vector spaces,
see, for example,~\cite{GJ,GK}),
i.e.~a sequence $\calp = \{\calp(n);\ n\geq 1\}$ of graded vector
spaces
together with degree zero linear maps
\[
\gamma=\gamma_{m_1,\ldots,m_l}:\calp(l)\ot\calp(m_1)\ot\cdots\ot%
\calp(m_l)
\longrightarrow
\calp(m_1+\cdots+m_l)
\]
given for any $m_1,\ldots,m_l \geq 1$. The usual axioms are assumed,
including the existence of an identity $1\in \calp(1)$. To stress the
existence of the unit $1$ we say that such an operad is {\em
unital\/}.

\begin{definition}
\label{operad_def}
A nonunital operad is a sequence $\cals =\{\cals(n);\
n\geq 1\}$ of graded vector spaces such that
\begin{enumerate}
\item[(i)]
Each $\cals(n)$ is equipped with a $\bk$-linear (left) action of the
group $\Sigma_n$, where $\Sigma_n$ is the symmetric group on $n$
elements, $n\geq 1$.
\item[(ii)]
We have, for any $m,n$ and $1\leq i\leq n$, degree zero linear maps
\[
\circ_i:\cals(m)\ot\cals(n)\to\cals(m+n-1)
\]
such that, for $f\in \cals(a)$, $g\in \cals(b)$ and $h\in \cals(c)$,
\begin{equation}
\label{1}
f\comp_i(g\comp_jh) =
\left\{
\begin{array}{ll}
(-1)^{|f|\cdot|g|}\cdot g\comp_{j+a-1}(f\comp_i h),&
1\leq i\leq j-1,
\\
(f \comp_{i-j+1}g)\comp_j h, &
j\leq i \leq b+j-1, \mbox{ and}
\\
(-1)^{|f|\cdot|g|}\cdot g\comp_j(f\comp_{i-j+1}h),&
i\geq j+b.
\end{array}
\right.
\end{equation}
\item[(iii)]
Let $\sigma \in \Sigma_m$, $\tau \in \Sigma_n$ and let
$\sigma\comp_i \tau \in \Sigma_{m+n-1}$ be given by
inserting the permutation $\sigma$ at the $i$th place in $\tau$.
The operations $\comp_i$ are equivariant in the sense that
\begin{equation}
\label{2}
(\sigma f)\comp_i (\tau g) = \sigma\comp_i \tau(f\comp_i g),
\end{equation}
for any $f\in \cals(m)$ and $g \in \cals(n)$.
\end{enumerate}
\end{definition}

Notice the resemblance of~(\ref{1})
with the definition of a comp algebra of~\cite{G}.
We say that a unital operad $\calp$ is {\em augmented\/} if there
exists a
homomorphism of unital operads $\epsilon : \calp \to \bk$ which
splits the
identity homomorphism $\eta :\bk \to \calp$ ($\bk$ denotes both the
ground
field and the trivial unital operad).

Let $\calp$ be an augmented operad and define
\[
\barcalp(n):=
\left\{
\begin{array}{ll}
\calp(n),&\mbox{for } n\geq 2, \mbox{ and}
\\
\Ker(\epsilon(1):\calp(1)\to \bk(1)=\bk),& \mbox{for $n=1$}.
\end{array}
\right.
\]
For $f\in \barcalp(m)$ and $g\in \barcalp(n)$ let
\[
f\comp_i g := \gamma_{1,\ldots,m,\ldots,1}
(g\ot1\cdots\ot f\ot\cdots\ot1)\in \barcalp(m+n-1)
\mbox{ ($f$ at the $i$th place)},
\]
$\gamma$ being the structure map of $\calp$. It is easy to see that
this
gives on $\barcalp$ the structure of a nonunital operad.

On the other hand, for a nonunital
operad $\cals$ set
\[
\hatcals(n):=
\left\{
\begin{array}{ll}
\cals(n),&\mbox{for } n\geq 2, \mbox{ and}
\\
\cals(1)\oplus\bk,& \mbox{for $n=1$}.
\end{array}
\right.
\]
Extend the structure maps $\comp_i
:\cals(m)\ot\cals(n)\to \cals(m+n-1)$ to $\hat\comp_i
:\hatcals(m)\ot\hatcals(n)\to \hatcals(m+n-1)$ by
\[
f\hat\comp_1 1 := f \mbox{ and } 1\hat\comp_i g := g
\]
for $f\in \hatcals(m)$, $g\in \hatcals(n)$, $m,n\geq 2$ and
$1\leq i\leq n$. Then the formula
\[
\gamma_{m_1,\ldots,m_l}(\mu\ot\nu_1\ot\cdots\ot\nu_l):=
\nu_1\hat\comp_1(\nu_2\hat\comp_2
(\cdots\hat\comp_{n-1}(\nu_n\hat\comp_n\mu)\!\cdot\!\!\cdot\!))
\]
defines on $\cals$ the structure of an unital operad.
Notice that an unital operad $\calp$ with $\calp(1)=\bk$ is always
canonically augmented, by defining $\epsilon(1):=\id$, the identity
map. We may formulate the following observation.

\begin{observation}
\label{equivalence}
The correspondence above defines an equivalence between the category
of
unital augmented operads and the category of nonunital operads. This
equivalence restricts to an equivalence of the category of unital
operads
$\calp$ with $\calp(1)=\bk$ and the category of
nonunital operads $\cals$ with $\cals(1)=0$.
\end{observation}

\noindent{\bf Warning.}
Notice that nonunital operads are {\em not\/} the same as the objects
whose axioms are obtained from the axioms for unital operads by
forgetting everything related with the unit! If $\calp$ is such an
object, there is no way of defining the operations $\comp_i$ from the
structure maps of~$\calp$.
\smallskip

In what follows we will be primarily concerned with
nonunital operads.
The advantage of working with nonunital operads is that their axioms
are quadratic (hence homogeneous) which will give a
nice natural filtration on the corresponding free objects, see
below. On the other hand, as we have seen in
Observation~\ref{equivalence}, nonunital operads are the same as
augmented
unital operads, therefore the results and constructions
of~\cite{GK,GJ} are in fact
available also for nonunital operads.

The main results of the paper will be, unless otherwise stated,
formulated for nonunital operads
\S\ with $\cals(1)=0$ (which are the same as unital operads $\calp$
with $\calp(1)=\bk$), with one very important exception --
topological and related singular chain operads of par.~5.
We also assume that all operads
in the paper are of {\em finite type\/}, meaning that $\cals(n)$ is a
finite dimensional vector space or, if $\cals$ is a differential
operad (see below for the definition),
$\cals = (\cals,d)$, that the cohomology
$H(\cals(n),d(n))$ is a finite dimensional vector space, for any
$n\geq 2$. The last assumption is not really necessary everywhere,
but it will simplify the exposition.

The following two constructions will be useful in the sequel. For an
operad $\cals$ define its {\em suspension\/} $\ssusp\cals$ to be the
operad defined by
\[
(\ssusp\cals)(n) = \susp^{n-1} \cals(n),
\mbox{ the $(n-1)$-fold suspension of the graded vector space
$\cals(n)$,}
\]
with the structure maps and the symmetric group action
defined in the obvious way. For a nonunital
operad $\calt$, {\em not\/} necessarily with $\calt(1) = 0$,
define its {\em
universal covering\/} $\widetilde\calt$ as the operad with
\[
\widetilde\calt(n):=
\left\{
\begin{array}{ll}
\calt(n),&\mbox{for } n\geq 2, \mbox{ and}
\\
0,& \mbox{for $n=1$},
\end{array}
\right.
\]
and the operad structure defined in the most obvious way. Using the
terminology introduced later, we may say alternatively that
$\widetilde\calt$ is the ideal in $\calt$ generated by $\{\calt(n);\
n\geq 2\}$.

By a {\em collection\/} we mean a sequence $E = \{E(n);\ n\geq 2\}$
of (graded) vector spaces such that each $E(n)$ is equipped with an
action of the symmetric group $\Sigma_n$. If $F = \{F(n);\ n\geq 2\}$
is another collection, then by a {\em map\/}
$\alpha: E \to F$ of collections we mean a sequence
$\alpha = \{\alpha(n);\ n\geq 2\}$ of $\Sigma_n$-equivariant maps
$\alpha(n): E(n)\to F(n)$. Instead of saying ``$\alpha$ is a map of
collections'' we sometimes say simply ``$\alpha$ is an equivariant
map''
hoping it will not confuse the reader.

Denote by $\Box :\oper \to
\coll$ the obvious forgetful functor from the category of
(nonunital) operads to the category of collections.
It has a left adjoint $\free
:\coll \to \oper$ (see~\cite{GJ,GK}) and it is natural to call
$\free(E)$ the {\em free operad\/} on the collection $E$. It follows
easily from the homogeneity of the axioms of an (nonunital) operad
that each $\free(E)(n)$ is naturally graded,
$\free(E)(n)=\bigoplus_{l\geq 1}\free^l(E)(n)$ and that the grading
has the following properties.
\begin{enumerate}
\item[(i)]
Each $\free^l(E)(n)$ is a $\Sigma_n$-invariant subspace of
$\free(E)(n)$.
\item[(ii)]
$\free^1(E)(n) = E(n)$ and $\free^{\geq n}(E)(n)=0$.
\item[(ii)]
If $f\in \free^l(E)(m)$ and $g\in \free^k(E)(n)$ then $f\comp_i g\in
\free^{k+l}(E)(m+n-1)$.
\end{enumerate}
In the tree language of~\cite{GK}, $\free^l(E)(n)$ is represented by
trees having $(l-1)$ inner edges.

Let $\cals$ be an operad. Define the collection $Q =Q(\cals)$ (the
{\em
indecomposables\/} of $\cals$) by $Q(n) := \cals(n)/D(n)$, where the
collection $D=D(\cals)$
(the {\em decomposables\/} of $\cals$) is defined as
the subcollection of $\cals$ spanned by the elements of
the form $\sigma(f\comp_i g)$ with $\sigma \in \Sigma_{m+n-1}$,
$m,n\geq 1$, $f\in \cals(m)$, $g\in \cals(n)$ and
$1\leq i\leq n$. We have the natural projection $\pi: \cals \to Q$ of
collections. Notice that there always exists an equivariant splitting
(i.e.~a map of collections) $s:Q\to \cals$ of $\pi$. To see this,
choose, for $n\geq 2$, a $\bk$-linear splitting $s'(n):Q(n)\to
\cals(n)$ of $\pi(n)$ and let $s(n)$ be the symmetrization
\begin{equation}
\label{symmetrization}
s(n):= \frac1{n!}\sum_{g\in\Sigma_n} gs'(n)g^{-1}.
\end{equation}
Then $s(n)$ is obviously an $\Sigma_n$-{\em equivariant\/} splitting
of
$\pi(n)$ and $s := \{s(n);n\geq 2\}$ is the required section. We will
use this trick in many places of the paper.

The following proposition shows that, as far as the presentation is
concerned, operads behave as connected graded algebras
(compare~\cite{ML}).

\begin{proposition}
\label{po}
In the situation above, define $\alpha :\free(Q) \to \cals$ by
$\alpha |_Q = s$. Then:
\begin{enumerate}
\item[(i)]
The map $\alpha$ is an epimorphism (meaning that each $\alpha(n)$ is
epi).
\item[(ii)]
$\Ker(\alpha)$ consists of reducible elements, $\Ker(\alpha)\subset
\free^{\geq 2}(Q)= D(\free(Q))$.
\item[(iii)]
The presentation $\alpha :\free(Q) \to \cals$ is minimal in the sense
that for any collection $Q'$ and for any
epimorphism $\alpha' :\free(Q') \to \cals$ there exists
a monomorphism $\beta :\free(Q) \to \free(Q')$ such that the diagram
\[
\hskip1.5cm
\Ytriangle{\free(Q)}{\cals}{\free(Q')}{\beta}{\alpha'}{\alpha}
\]
commutes.
\end{enumerate}
\end{proposition}

\noindent{\bf Proof.}
To prove that $\alpha$ is an epimorphism,
observe first that $Q(2)=\cals(2)$,
hence $\alpha(2)=\id$. Suppose we have proved that
$\alpha(k)$ is epi for any $2\leq k< n$ and try to prove that
$\alpha(n)$
is an epimorphism too. Any $f\in \cals(n)$ can be
decomposed uniquely as $f= s(n)(x) + y$ for some $x\in Q(n)$ and
$y\in
\cals(n)$ with $\pi(n)(y)=0$. We have, by definition,
$s(n)(x)=\alpha(n)(x)$, so it is enough to prove that $y\in
\mbox{Im}(\alpha(n))$. But $\pi(n)(y)=0$ means that $y\in D(n)$,
i.e.
\[
y= \sum_t \sigma_t(f_t\comp_{i_t}g_t)
\]
for some $\sigma_t\in \Sigma_n$, $f_t\in \cals(m_t)$, $g_t\in
\cals(n_t)$
and $1\leq i_t\leq n_t$. We know by induction (since $m_t,n_t
< n$) that $f_t=\alpha(m_t)(\phi_t)$ and $g_t = \alpha(n_t)(\psi_t)$
for
some $\phi_t\in \free(Q)(m_t)$ and $\psi_t\in \free(Q)(n_t)$. As
$\alpha$
is a homomorphism of operads we have
\[
y= \alpha(m+n-1)(\sum_t \sigma_t(\phi_t\comp_{i_t}\psi_t)),
\]
hence $y\in \mbox{Im}(\alpha(n))$ and (i) is proven by induction.

To prove (ii), observe first that $\alpha$ maps $\free^{\geq 2}(Q)$
into
the decomposables $D$ of $\cals$. Let $f\in \Ker(\alpha(n))\subset
\free(Q)(n)$ for some $n\geq 2$. We may decompose $f=f_1+f_{\geq 2}$
with $f\in
\free^1(Q)(n) = Q(n)$ and $f\in \free^{\geq 2}(Q)(n)$. Then we have
\[
0=\pi(n)\alpha(n)(f)= \pi(n)s(n)(f_1)+\pi(n)\alpha(n)(f_{\geq 2}) =
f_1,
\]
hence $f\in \free^{\geq 2}(Q)(n)$ as claimed.

We leave (iii) as an exercise for the reader.
\qed

The notions as (co)kernels, ideals, quotients, etc., translate in the
obvious way into the category $\oper$. For example, an {\em ideal\/}
in
$\cals$ is a suboperad $\cali = \{\cali(n)\subset \cals(n);\ n\geq
2\}$ such that $f\comp_i g\in I(m+n-1)$ whenever $f\in \cali(m)$ or
$g\in
\cali(n)$, $m,n\geq 1$, $1\leq i\leq n$. If $\cali$ is an ideal in
$\cals$ then
the quotient $\cals/\cali:=\{\cals(n)/\cali(n);\ n\geq 2\}$ has a
natural structure of an operad,~etc.

Let $\cals$ be an operad. As usual, we have several equivalent
definitions
of a module over $\cals$; for example, we can simply say that a
module is
an abelian group object in the slice category $\oper/\cals$ of
operads over
$\cals$. But we prefer to give the following more explicit
definition,
compare the notion of a module over a {\em prop\/} introduced
in~\cite{M}.

\begin{definition}
\label{module}
A module over an operad $\cals$ is a collection $M=\{M(n);\ n\geq
2\}$
together with degree zero linear maps
\begin{eqnarray*}
\comp_i=\comp_i^L &:& \cals(m)\ot M(n) \to M(m+n-1) \mbox{ (the left
composition), and}
\\
\comp_i=\comp_i^R &:& M(m)\ot \cals(n) \to M(m+n-1) \mbox{ (the right
composition),}
\end{eqnarray*}
given for any $m,n\geq 2$ and $1\leq i\leq n$
such that the condition~(\ref{1}) of Definition~\ref{operad_def} is
satisfied for any
$f\ot g\ot h \in [M(a)\ot\cals(b)\ot\cals(c)] \mbox{ or }
[\cals(a)\ot M(b)\ot\cals(c)]\mbox{ or }
[\cals(a)\ot\cals(b)\ot M(c)]$
and that the condition~(\ref{2}) of Definition~\ref{operad_def} is
satisfied for any $f\ot g \in [M(m)\ot \cals(n)]\mbox{ or }
[\cals(m)\ot
M(n)]$.
\end{definition}

\begin{example}{\rm\
\label{modules}
There are two important types of examples of \S-modules. If $\phi
:\cals \to\calt$ is a homomorphism of operads, then $\phi$ induces on
$\calt$ the natural structure of an operad.
In particular (taking $\phi = \id$),
\S\ is an \S-module over itself.

Recall that an {\em algebra over \S\/} (or an \S-{\em algebra\/})
is a homomorphism
$A:\cals \to \End(V)$, where $\End(V)$ denotes the operad of
endomorphisms
of a graded space $V$ (see~\cite{GJ,GK}, compare also the differential
case discussed later). Therefore $A$ induces on
$\End(V)$ the structure of an \S-module which we denote (a bit
ambiguously) by $\bk_V$, because this module plays the r\^ole of the
residue ring in our theory, see also the discussion in~\cite{M}.

A second example is provided by
an ideal $K\subset \cals$, typically given as the
kernel of a homomorphism. The operad structure of $\cals$ obviously
induces
on $K$ the natural structure of an \S-module.
}\end{example}

The forgetful functor $\Box :\smod \to \coll$ from the category of
$\cals$-modules to the category of collections has a left adjoint
(compare~\cite{M}), $\frees-:\coll \to \smod$, and we call the
$\cals$-module
$\frees E$ the {\em free $\cals$-module\/} on the collection $E$. The
free
module $\frees E$ is, like the free operad, naturally graded,
$\frees E = \bigoplus_{l\geq 1}{\frees E}^l$, by the ``length of
words''.

For an $\cals$-module $M$ define the collection $Q_\cals =Q_\cals(M)$
(the
{\em indecomposables\/} of $M$) by $Q_\cals(n) := M(n)/D_\cals(n)$,
where the
collection $D_\cals=D_\cals(M)$
(the {\em decomposables\/} of $M$) is defined as
the subcollection of $M$ spanned by the elements of
the form $\sigma(f\comp_i g)$ with $\sigma \in \Sigma_{m+n-1}$,
$m,n\geq 1$, $f\in M(m)$, $f\ot g \in M(m)\ot \cals(n)\oplus
\cals(m)\ot
M(n)$ and
$1\leq i\leq n$. We have the following analog of
Proposition~\ref{po}.

\begin{proposition}
\label{pm}
Let $M$ be an $\cals$-module and let $t:Q_\cals \to M$ be a section of
the
natural projection $p :M \to Q_\cals$. Define $\gamma
:\frees{Q_\cals}\to
M$ by $\gamma|_{Q_\cals} = t$. Then the following holds.
\begin{enumerate}
\item[(i)]
The map $\gamma$ is an epimorphism.
\item[(ii)]
$\Ker(\gamma)$ consists of decomposable elements, $\Ker(\gamma)\subset
\frees{Q_\cals}^{\geq 2}= D_\cals(\frees{Q_\cals})$.
\item[(iii)]
The presentation $\gamma :\frees{Q_\cals} \to M$ is minimal in the
sense
that for any collection $Q$ and for any
epimorphism $\gamma' :\frees{Q} \to M$ there exists
a monomorphism $\delta :\frees{Q_\cals} \to\frees{Q}$ such that the
diagram
\[
\hskip1.5cm
\Ytriangle{\frees{Q_\cals}}{\cals}{\frees{Q}}{\delta}{\gamma'}{\gamma}
\]
commutes.
\end{enumerate}
\end{proposition}

\noindent{\bf Proof.}
It is an obvious analog of the proof of Proposition~\ref{po}
and we can leave it to the reader. \qed

By a {\em degree $p$ derivation\/} on an operad $\cals$ we mean a
sequence $\theta
=\{\theta(n):\cals(n)\to \cals(n);\ n\geq 2\}$ of equivariant degree
$p$
maps such that
\[
\theta(m+n-1)(f\comp_i g) =
\theta(m)(f)\comp_i g + (-1)^{p\cdot|f|}\cdot f\comp_i \theta(n)(g),
\]
for any $f\in \cals(m)$, $g\in \cals(n)$, $m,n\geq 1$ and $1\leq i\leq
n$.
We denote by $\der_p(\cals)$ the vector space of all degree $p$
derivations of $\cals$.

\begin{lemma}
Let $\phi \in \der_p(\cals)$ and $\psi \in \der_q(\cals)$. Then the
formula
\[
[\phi,\psi](n) := \phi(n)\psi(n) - (-1)^{pq}\psi(n)\phi(n),\ n\geq
2.
\]
defines on $\der_*(\cals)$ the structure of a graded Lie algebra.
\end{lemma}

\noindent{\bf Proof.}
An easy exercise. \qed

By a {\em differential operad\/} we mean a couple $(\cals,d)$ where
$\cals$
is an operad and $d$ is a degree $-1$ derivation of $\cals$, $d \in
\der_{-1}(\cals)$, with $d^2 = 0$. It is clear that for such a
differential
operad the collection $\calh = \calh(\cals,d) :=
\{H(\cals(n),d(n));\ n\geq
2\}$, where $H(\cals(n),d(n))$ denotes the homology of the
differential
space $(\cals(n),d(n))$, has a natural structure of an operad, called
the {\em homology (operad)\/} of $(\cals,d)$. Observe that what we
call
a differential operad here
is exactly an operad in the monoidal category of
differential spaces.

Of course, any nondifferential operad $\cals$ can be considered as a
differential operad with the trivial differential. If we wish to
stress that we consider $\cals$ in this way, we write
$(\cals,0)$ instead of $\cals$.

\begin{example}{\rm\
\label{end}
Any graded differential space $(V,d_V)$ has its (nonunital) {\em
endomorphism operad\/} $(\End(V),\dend)$ defined by
\[
\End(V)(n):=
\left\{
\begin{array}{ll}
\mbox{Hom}_*(V^{\ot n},V),&\mbox{for } n\geq 2, \mbox{ and}
\\
0,& \mbox{for $n=1$},
\end{array}
\right.
\]
where $\mbox{Hom}_p(V^{\ot n},V)$ denotes the vector space of
homogeneous
degree $p$ linear maps $f:V^{\ot n}\to V$. The composition maps are
defined in the obvious way and the differential $\dend$ is, for $f\in
\mbox{Hom}_p(V^{\ot n},V)$ given by
\[
\dend(f):= d\circ f -(-1)^p\cdot f\circ d^{\ot n},
\]
where $d^{\ot n}$ is the usual differential induced by $d$ on $V^{\ot
n}$.
An {\em algebra\/} over a differential operad $(\cals,d_\cals)$ (or an
$(\cals,d_\cals)$-{\em algebra}) is then a differential operad
homomorphism
$A:(\cals,d_\cals)\to (\End(V),\dend)$.
}\end{example}

For any two collections $X$ and $Y$, let $\mbox{\rm Coll}_p(X,Y)$
denotes the space of all sequences $\phi = \{\phi(n);\ n\geq 2\}$ of
$\Sigma_n$-equivariant degree $p$ linear maps $\phi(n):X(n)\to
Y(n)$.
Using the universal property of the free operad, we may easily deduce
the following lemma.

\begin{lemma}
\label{restriction}
For any collection $E$, the restriction defines an isomorphism
\[
\der_p(\free(E))\cong \mbox{\rm Coll}_p(E,\free(E)).
\]
\end{lemma}

Let us state also the following technical lemma.

\begin{lemma}
\label{micinka}
The correspondence $\calh:(\cals,d_\cals) \mapsto
\calh(\cals,d_\cals)$
induces a functor from the category of differential operads to the
category of nondifferential operads.

If $(V,d_V)$ is a differential vector space and $(\End(V),\dend)$ the
endomorphism operad from Example~\ref{end}, then there exists a
noncanonical map $\Phi : (\End(H(V,d_V),0) \to (\End(V),\dend)$ of
differential operads inducing the canonical isomorphism
\[
(\End(H(V,d_V),0)\cong \calh(\End(V),\dend)
\]
given by the K\"unneth formula.
\end{lemma}

\noindent{\bf Proof.}
The first part of the lemma is an easy exercise.
To prove the second part, choose a decomposition $V = H\op B \op C$
with
$H\op B = \Ker(d_V)$ and $B = \Im(d_V)$. Let $\iota :H \to V$ be the
corresponding inclusion and $\pi :V \to H$ be the corresponding
projection.
For an element $f \in \End(H)(n)$, $f: H^{\ot n} \to
H$, let $\Phi(f)\in
\End(V)$ be defined as $\Phi(f):= \iota \circ f(\pi^{\ot n})$. Let us
verify that $\Phi : (\End(H(V,d_V),0) \to (\End(V),\dend)$ thus
defined
is a homomorphism of differential operads.

For $f\in \End(H)(m)$, $g\in \End(H)(n)$ and $1\leq i\leq n$, we have
\begin{eqnarray*}
\Phi(f\comp_i g)& =& \iota\circ g(\id^{\ot(i-1)}\ot f\ot
\id^{\ot(n-i)})(\pi^{\ot(m+n-1)})
=\iota\circ g(\pi^{\ot(i-1)}\ot f(\pi^{\ot m})\ot \pi^{\ot(n-i)})
\\
& =& \iota\circ g(\pi^{\ot(i-1)}\ot\pi\iota
f(\pi^{\ot m})\ot \pi^{\ot(n-i)})
=\Phi(f)\comp_i\Phi(g),
\end{eqnarray*}
because $\pi\iota = \id$. We also easily have $\dend(\Phi(f)) = d
\iota f \pi^{\ot m} - (-1)^p \iota f (\pi d)^{\ot n}= 0$, because
$d\iota = \pi d =0$, thus $\Phi$ is indeed a map of differential
operads. The rest of the statement follows from the K\"unneth
formula.
\qed

Let $A:(\cals,d_\cals) \to (\End(V),\dend)$ be an
$(\cals,d_\cals)$-algebra structure on $(V,d_V)$. By the above lemma,
$A$ induces an $\calh(\cals,d_\cals)$-algebra structure on
$H(V,d_V)$,
\[
\calh(A) : \calh((\cals,d_\cals)\to \End(H(V,d_V)
\]
which we call the {\em homology algebra structure\/} induced by $A$.

All the objects introduced above and the majority of objects
introduced below have an obvious and even easier {\em nonsymmetric\/}
form which means that we simply forget everything related to the
action of the symmetric group, compare the similar situation in
homotopy theory~\cite{May}.

\section{Self-dual nature of the category of operads}

This section has an auxiliary character though we think that the
results may be of some interest in themselves. We aim to discuss
here the
following statement which is implicit in~\cite{GK,GJ}.

\begin{theorem}
\label{self-dual_nature}
Let $E$ be a (finite-type) collection. Then there exists a one-to-one
correspondence between (nonunital) operad structures on $E$ and
quadratic differentials on $\free(\#\desusp E)$ (i.e.~differentials
$d$
such that $d(\#\desusp E)\subset \free^{2}(\#\desusp E))$,
\[
\Omega:\{\mbox{operad structures on $E$}\}
\longleftrightarrow
\{\mbox{quadratic differentials on $\free(\#\desusp E)$}\}:\Omega^{-1}
\]
where $(\#\desusp E)$ is the collection
given by $(\#\desusp E)(n)= \#\desusp
E(n)$, $n\geq 2$. The functor $\Omega$ (the dual bar construction) was
explicitly constructed in~{\rm\cite{GK}}.
\end{theorem}

Let $\calr$ be a $\bk$-linear graded
rigid monoidal category, i.e.~a category
where both the objects and the hom-sets are given a structure of
$\bk$-linear graded vector space, a natural transformation
$\odot:\calr \times \calr \to \calr$ (the ``tensor product'') is
given, and every object $V\in \calr$ has a dual $\#V \in \calr$. We do
not aim to give a full set of axioms here; we just mention two major
examples instead: the category of graded vector spaces with the usual
tensor product and dual, and the category $\coll$ of collections with
the rigid monoidal structure defined below.

For a category $\calr$ as above and $V\in \calr$,
let $\End_\calr(V)$ be a generalization of the
nonsymmetric operad $\End(V)$ from
Example~\ref{end} with $\calr$ in place of $V$
defined by
\[
\End_\calr(V)(n):= \calr(V^{\odot n},V),\ n\geq 2,
\]
with the structure maps $\comp_i$ given as the obvious compositions.
If \S\ is an operad, then an \S-{\em algebra\/} in $\calr$ is a
homomorphism $A:\cals \to \End_\calr(V)$ of operads.

The following proposition is a generalized dual form of a theorem
of~\cite{FM}.

\begin{proposition}
\label{zebrulka}
Let $\calr$ be a $\bk$-linear graded rigid monoidal category and let
$V\in \calr$. Let $\cals$ be a quadratic operad and let $\cals^!$ be
its Koszul dual (see~{\rm\cite{GK}} for definitions). Then there is a
natural one-to-one correspondence between $\cals$-algebra structures
on a given object $V\in \calr$ and quadratic differentials on
$F_{\cals^!}(\#\desusp V)$, where $F_{\cals^!}(\#\desusp V)$ denotes
the
free $\cals^!$-algebra on $V$.
\end{proposition}

Let us introduce now a rigid monoidal structure on the category
of collections $\coll$.
For two collections $E,F \in \coll$, let $E\odot F$ be the
collection defined as
\[
\mbox{$
(E\odot F)(n):=\bigoplus_{k+l=n+1}
\bk[\Sigma_n] \left\langle
\bigoplus_{1\leq i\leq l}(E(k)\ot_i F(l))\right\rangle
/ ({\sim}),$}
\]
where $\bk[\Sigma_n]\left\langle \bigoplus_{1\leq i\leq l}
(E(k)\ot_i F(l)) \right\rangle$
is the free $\Sigma_n$-space on $l$ copies
of $(E(k)\ot_i F(l)):=(E(k)\ot F(l))$ and the equivalence relation
$\sim$ is generated by
\[
(\sigma\comp_i \tau)(e\ot_i f) \sim \sigma e \ot_i \tau f,
\]
the meaning of $\sigma\comp_i \tau\in \Sigma_n$
being the same as in (iii) of
Definition~\ref{operad_def}. The definition of the dual collection
$\#E$ is the obvious one.
The following lemma is a nice exercise left for the reader
(compare~\cite{S}).

\begin{lemma}
Nonunital operads are associative algebras in the category $\coll$
with the monoidal structure introduced above.
\end{lemma}

The philosophy behind Theorem~\ref{self-dual_nature}
becomes obvious now. An associative algebra is an algebra over the
associative algebra operad {\it Ass\/} which is,
by~\cite{GK}, Koszul self-dual, ${\it Ass\/}^! =
{\it Ass\/}$, and we see that
Theorem~\ref{self-dual_nature} is a consequence of
Proposition~\ref{zebrulka}.

\section{Models for operads}

In the following theorem, we show that there exists an analog of
a minimal model for differential operads.

\begin{theorem}
\label{minimal}
Let $(\cals,d_\cals)$ be a differential graded operad (with
$\cals(1)=0$ as
usual). Then there exists a collection $M=\{M(n);\ n\geq 2\}$, a
differential $d$ on $\free(M)$ and a homomorphism $\nu
:(\free(M),d)\to
(\cals,d_\cals)$ such that
\begin{enumerate}
\item[(i)]
the differential $d$ is minimal, $d(M)\subset \free^{\geq 2}(M)$, and
\item[(ii)]
the map $\nu$ induces an isomorphism in homology.
\end{enumerate}
The object $\nu :(\free(M),d)\to
(\cals,d_\cals)$ is called the minimal model of the differential
operad
$(\cals,d_\cals)$. It is unique in the sense that if $\nu'
:(\free(M'),d)\to
(\cals,d_\cals)$ is another minimal model of $(\cals,d_\cals)$, then
the differential operads $(\free(M),d)$ and $(\free(M'),d)$ are
isomorphic.
\end{theorem}

\noindent{\bf Proof.}
Let $X(2):= \calh(\cals,d_\cals)(2)$ and let $s(2):X(2)\to Z
(\cals,d_\cals)(2)\subset \cals(2)$ be an
equivariant splitting of the
projection $\cl(2): Z (\cals,d_\cals)(2)\to
\calh(\cals,d_\cals)(2)$ (here and later, $\cl$ will denote the
projection of a chain onto its homology class).
Define a differential $d_2$ on $\free(X(2))$
and a map $\nu_2:(\free(X(2),d_2)\to (\cals,d_\cals)$ by
\[
d_2=0 \mbox{ and } \nu_2|_{X(2)}:=s(2).
\]

We finish the construction by induction. Suppose we have already
constructed a collection
$X(<\!\! n) = \{X(k);\ 2\leq k< n\}$, a differential $d_{n-1}$
on $\free(X(<\!\! n))$ and a map $\nu_{n-1}:(\free(X(<\!\!
n)),d_{n-1})\to
(\cals,d_\cals)$
such that
\begin{center}
\begin{minipage}{16cm}
\baselineskip18pt
\begin{enumerate}
\item[$\mbox{(i)}_{n-1}$]
the differential $d_{n-1}$ is minimal, $d_{n-1}(X(<\!\! n))\subset
\free^{\geq 2}(X(<\!\! n))$ and
\item[$\mbox{(ii)}_{n-1}$]
the map $\calh(\nu_{n-1})(k): \calh(\free(X(<\!\! n)),d_{n-1})(k)
\to (\cals,d_\cals)(k)$ is an isomorphism for any $k\leq n-1$.
\end{enumerate}
\end{minipage}
\end{center}
Let
\[
A(n):=
\calh (\cals,d_\cals)(n)/(\Im(\calh(\nu_{n-1}))(n),\
\overline B(n):=
\Ker(\calh(\nu_{n-1}))(n)
\mbox{ and } B(n):= \susp \overline B(n).
\]
Let $s(n):A(n)\to Z (\cals,d_\cals)(n)$ be an equivariant section of
the
composition $ Z(\cals,d_\cals)(n)\overto{\cl} A(n)$ and let
$r'(n): \calh(\free(X(<\!\! n)),d_{n-1})(n)\to
Z(\free(X(<\!\! n)),d_{n-1})(n)$ be an
equivariant section of the projection
$\cl_n:Z(\free(X(<\!\! n)),d_{n-1})(n)\to
\calh(\free(X(<\!\! n)),d_{n-1})(n)$ and let
$r(n):\overline B(n)\to
Z(\free(X(<\!\! n)),d_{n-1})(n)$ be the composition of the inclusion
$\overline B(n)\hookrightarrow
\calh(\free(X(<\!\! n)),d_{n-1})(n)$ and $r'(n)$.
Define
\begin{eqnarray*}
&&X(n) := A(n)\op B(n),\ X(\leq \!\! n):= X(<\!\! n)\op X(n),
\\
&&d_n|_{X(< n)} := d_{n-1}|_{X(< n)},\
d_n|_{A(n)}:= 0,\ d_n|_{B(n)}:=
r(n)\circ\desusp~,
\\
&&\nu_n|_{X(< n)} := \nu_{n-1}|_{X(< n)},\
\nu_n|_{A(n)}:= s(n),\mbox{ and }
\nu_n|_{B(n)}:= 0.
\end{eqnarray*}
Then $d_n$ is obviously minimal since
$Z(\free(X(<\!\! n)),d_{n-1})(n)\subset
\free(X(<\!\! n))(n)\subset
\free^{\geq 2}(X(<\!\! n))(n)$. Also $\calh(\nu_n)(n)$ is
obviously an epimorphism, by construction.
Let us prove that $\calh(\nu_n)(n)$ is a monomorphism.

Let $z\in Z(\free(X(\leq n)),d_n)(n)$ and write it in the form $z=a+
b+ \omega$, with $a\in A(n)$,
$b\in B(n)$ and $\omega\in \free^{\geq
2}(X(<\!\! n))$. By definition $d_n(a)=0$,
therefore $-d_n(b)=-r(n)(\desusp b)=d_n(\omega)$.
Since $\omega$ is decomposable we have, in fact, $d_n(\omega) =
d_{n-1}(\omega)$, therefore $d_n(\omega)$ represents a trivial
homological class in $\calh(\free(X(<\!\! n)),d_{n-1})(n)$ and $b=0$.
If $z$ represents an element of $\Ker(\calh(\nu_n))(n)$, then $a=0$,
so $z\in Z(\free(X(<\!\! n)),d_{n-1})(n)$ and obviously $\cl(z)\in
\Ker(\calh(\nu_{n-1}))(n)$. But then $z=d_n(\susp \cl_n(z))$
by the construction of $d_n$ which implies that $z$ represents a
trivial homology class in $\calh(\free(X(\leq\!\! n)),d_n)$.
We see that our data satisfy the conditions
$\mbox{(i)}_{n}$--$\mbox{(ii)}_{n}$ and we finish the construction
by induction.

We leave the proof of the uniqueness to the reader.
\qed

Let $Y$ be a collection and $\delta$ a (not necessary minimal)
differential on
$\free(Y)$. Let $\proj^i:\free(Y)\to
\free^i(Y)$ be the natural projection and
let $\delta_i$ be the derivation on
$\free(Y)$ defined by $\delta_i|_Y :=
\proj^i \circ \delta$. Notice that
\[
\delta_1^2 = 0 \mbox{ and } \delta_1\circ
\delta_2+\delta_2\circ\delta_1 = 0.
\]
This means that it makes sense to consider the
collection $H(Y,\delta_1)$ and
that $\delta_2$ induces on $\free(H(Y,\delta_1))$ a
quadratic differential
$\overline\delta_2$. There is an obvious important
special case when $\delta$
is minimal, which is the same as $\delta_1=0$.
Then $\overline\delta_2 =
\delta^2$ is a quadratic differential on $\free(Y)$.

The following definition is motivated by the notion of the homotopy
Lie
algebra in rational homotopy theory.

\begin{definition}
\label{homotopy_Lie}
Let $(\cals,d_\cals)$ be a differential operad
and let $\nu: (\free(M),d)\to
(\cals,d_\cals)$ be its minimal model as in
Theorem~\ref{minimal}. The
(nondifferential) operad $\Omega^{-1}(\free(M),d_2)$
(see Theorem~\ref{self-dual_nature} for the notation) is
called the homotopy
operad of $(\cals,d_\cals)$ and denoted $\pi(\cals,d_\cals)$.
\end{definition}

We state without proof, which is analogous to the proof
of~\cite[Theorem V.7]{L}, the following statement.

\begin{proposition}
\label{kralicek}
Let $(\free(Y),\delta)$ be as above and let $\nu:
(\free(M),d)\to(\free(Y),\delta)$ be its minimal model. Then
the differential operads $(\free(M),d_2)$ and
$(\free(H(Y,\delta_1)),\overline\delta_2)$ are isomorphic,
\[
(\free(M),d_2)\cong (\free(H(Y,\delta_1)),\overline\delta_2).
\]
\end{proposition}

Let us make some more comments on the dual
bar construction $\Omega$
of~\cite{GK}. It is defined for any differential
operad $(\cals,d_\cals)$ and
it is of the form $(\free(Y),d_\Omega)$,
with $Y:= \#\desusp \cals$ and
$d_\Omega := d_I + d_E$, where $d_I$
(the internal differential) is the dual
of $d_\cals$, and $d_E$ (the external
differential) is obtained by dualizing
the structure maps of $\cals$.

\begin{theorem}
Let $(\cals,d_\cals)$ be a differential operad. Then
\[
\pi(\cals,d_\cals) \cong \calh(\Omega(\cals,d_\cals)).
\]
\end{theorem}

\noindent{\bf Proof.}
By a theorem of~\cite{GK} there exists a natural map $\phi:
\Omega(\Omega(\cals,d_\cals))\to (\cals,d_\cals)$ such
that $\calh(\phi)$ is an
isomorphism. This means that, if $(\free(M),d)$ is a minimal model for
the double dual bar construction
$\Omega(\Omega(\cals,d_\cals))$, then it is also a minimal model for
$(\cals,d_\cals)$. We have
\[
\Omega(\Omega(\cals,d_\cals))=
(\free(\#\desusp \Omega(\cals,d_\cals)),d_\Omega),
\]
where $(d_\Omega)_1 = d_I$ is the dual of the differential on
$\Omega(\cals,d_\cals)$, therefore
\[
\free(H(\#\desusp\Omega(\cals,d_\cals),
(d_\Omega)_1),\overline\delta_2) =
\free(\#\desusp\calh(\Omega(\cals,d_\cals)),\overline\delta_2) =
\Omega(\calh(\Omega(\cals,d_\cals))),
\]
and Proposition~\ref{kralicek} gives the desired result.
\qed

Suppose that we have a collection $Z$ which decomposes as $Z=Z^0\op
Z^1\op\cdots$ (meaning, of course, that for each $n\geq 2$ we have a
$\Sigma_n$-invariant decomposition $Z(n) = Z^0(n)\op
Z^1(n)\op\cdots$).
This induces on $\free(Z)$ still another grading (because of the
homogeneity of the axioms of a nonunital operad!), $\free(Z) =
\bigoplus_{k\geq 1}\free(Z)^k$. We will call this grading the
TJ-grading
(from Tate-Jozefiak) here; the reason will became obvious below.

In this situation the free operad $\free(Z)$ is naturally trigraded,
$\free(Z) = \bigoplus \free^l(Z)^k_j$, where the $k$ refers to the
TJ-grading introduced above, the $l$ refers to
the ``length'' grading given by the length of words in the free
operad, and
the $j$ indicates the ``inner'' grading given by the grading of the
underlying vector space.

Suppose that $d$ is an (inner) degree $-1$ differential on $\free
(Z)$ such
that
\begin{equation}
\label{3}
d(Z^k) \subset \free(Z)^{k-1}, \mbox{ for all $k\geq 1$}
\end{equation}
(meaning, of course, that $d(Z^k)(n) \subset \free(Z)^{k-1}(n)$ for
all
$n\geq 2$), i.e.~that $d$ is homogeneous degree $-1$ with respect to
the
TJ-grading. Then the homology operad $\calh( \free (Z),d)$ is
naturally
bigraded,
\[
\calh( \free (Z),d) = \bigoplus \calh^k_j( \free (Z),d),
\]
the upper grading being induced by the TJ-grading and the lower one by
the inner grading.
We have the following analog of the bigraded model of a commutative
graded algebra constructed in~\cite[par.~3]{HS}.

\begin{theorem}
\label{TJ}
Let $\calh$ be a (nonunital with $\calh(1)=0$ as usual) operad. Then
there
exists a collection $Z=Z^0\op
Z^1\op\cdots$, a differential $d$ on $ \free (Z)$
satisfying~(\ref{3}) and
a map $\rho : (\free(Z),d)\to (\calh,0)$ of differential operads
such that the following conditions
are
satisfied:
\begin{enumerate}
\item[(i)]
$d$ is minimal in the sense that $d(Z)\subset
\free^{\geq 2}(Z)$,
\item[(ii)]
$\rho|_{Z^{\geq 1}}=0$ and $\rho$ induces an isomorphism
$\calh^0(\rho):\calh^0(\free(Z),d) \cong \calh$, and
\item[(iii)]
$\calh^{\geq 1}(\free(Z),d)=0$.
\end{enumerate}
We call $\rho :(\free(Z),d)\to(\calh,0)$ the TJ-model (or the
bigraded model) of $\calh$.
\end{theorem}

\noindent{\bf Proof.}
Let $Z^0:= Q(\calh)$ and let $s^0 : Z^0 \to \calh$ be an
(equivariant)
section of the projection $\calh \to Z^0$.
Define $\overline\rho : \free(Z^0)\to
\calh$ by $\overline\rho|_{Z^0} = s$ and let $d$ be trivial on
$\free(Z^0)$. Then $\calh^0(\overline\rho) = \overline\rho:
\free(Z^0) \to \calh$ is
an epimorphism by Proposition~\ref{po}.

Let $K := \Ker(\overline\rho)$ and consider the collection $K$ with
the
natural
structure of an $\free(Z^0)$-module as in Example~\ref{modules}.
Define
\[
\overline Z^1 := Q_{\free(Z^0)}(K)
\mbox{ and }
Z^1 = \susp \overline Z^1.
\]
Let $s^1 :\overline Z^1 \to K$ be a splitting of the projection $K
\to
\overline Z^1$ and extend $d$ by $d|_{Z^1} := s^1 \comp \desusp$~.
By Proposition~\ref{pm}, $K\subset \free^{\geq 2}(Z^0)$, therefore
$d(Z^1)\subset \free^{\geq 2}(Z^0)$. We have obviously
\[
\calh^0(\free(Z^{\leq 1}),d) = \free(Z^0)/K \cong \calh.
\]
To finish the proof, we construct inductively, for any $n\geq 2$, a
collection $Z^{<n} = Z^0\op Z^1\op \cdots \op Z^{n-1}$ (where $Z^0$
and
$Z^1$ are those already constructed above), a differential $d$ on
$\free(Z^{< n})$ (which extends that already constructed above) such
that, if $\rho :\free(Z^{< n}) \to \calh$ is given by
$\rho|_{Z^0}: = \overline\rho$ and $\rho|_{Z^k}:=0$ for $1\leq k< n$,
the
following conditions are satisfied:
\begin{center}
\begin{minipage}{16cm}
\baselineskip18pt
\begin{enumerate}
\item[$\mbox{(i)}_{n-1}$]
$d(Z^{<n})\subset \free^{\geq 2}(Z^{< n})$,
\item[$\mbox{(ii)}_{n-1}$]
The map $\rho$ induces an isomorphism $\calh^0(\rho):
\calh^0(\free(Z^{< n}),d)\cong
\calh$, and
\item[$\mbox{(iii)}_{n-1}$]
$\calh^k(\free(Z^{< n}),d)=0$ for any $1\leq k\leq n-2$.
\end{enumerate}
\end{minipage}
\end{center}
For $n=2$ these data have already been constructed above. Suppose we
have
constructed them for some $n\geq 2$.
The inclusion
$\free(Z^{0}) \hookrightarrow \free(Z^{<n})$ induces on
$\free(Z^{<n})$ the structure of an $\free(Z^{0})$-module which in
turn
induces on $\calh(\free(Z^{<n}),d)$ the structure of an
$\calh$-module (since $\calh^0(\free(Z^{<n}),d)=\calh$). Let
\[
\overline Z^n := Q_\calh(\calh^{n-1}(\free(Z^{<n},d)))
\mbox{ and }
Z^n:= \susp \overline Z^n.
\]
Let $s^n : \overline Z^n \to Z^{n-1}(\free(Z^{<n}),d))
\subset \free(Z^{< n})$ be an equivariant section of the composition
of the projections
\[
Z^{n-1}(\free(Z^{<n}),d))\overto{\cl}
\calh^{n-1}(\free(Z^{<n}),d))\to
Q_\calh(\calh^{n-1}(\free(Z^{<n}),d)).
\]
Extend the differential $d$ by $d|_{E^n}:= s^n\circ \desusp$~~and
prove that
the data thus constructed satisfy the conditions
$\mbox{(i)}_{n}$--$\mbox{(iii)}_{n}$. We show first that $d^2=0$. By
Lemma~\ref{restriction} it is enough to verify that $d^2=0$ on
generators. We have $d^2|_{Z^{<n}}=0$ by the induction
and $d^2|_{Z^{n}}=0$ because $d|_{Z^n}=s^n\circ \desusp$~~by
definition and
$\mbox{\rm Im}(s^n)\subset Z^{n-1}(\free (Z^{<n}),d)$.

To prove the minimality $\mbox{(i)}_{n}$ we must show that
$Z^{n-1}(\free(Z^{<n}),d) \subset \free^{\geq 2}(Z^{<n})^{n-1}$.
Any $u\in Z^{n-1}(\free(Z^{<n}),d)$
obviously decomposes as $u=z+w$ with
$z\in \free(Z^0\op Z^{n-1})$ and $w\in \free(Z^{<n-1})$. Further
analysis on
the TJ-grading shows that $z=z'+z''$, with $z'\in Z^{n-1}$ and $z''\in
D_{\free(Z^0)}( \free(Z^0\op Z^{n-1})$ while in fact $w\in \free^{\geq
2}(Z^{<n-1})$.
We infer from $0=dz = dz'+dz''+dw$ that $dz'$ represents an element
of $Q_\calh(\calh^{n-2}(\free(Z^{<n-1}),d)))$, therefore $z'=0$ by the
definition of $dz'$ as $s^{n-1}\circ \desusp z'$,
which means that $u$ is decomposable.

The condition $\mbox{(ii)}_{n}$ is obviously satisfied because our
construction does not change $\calh^0$ while $\mbox{(iii)}_{n}$
is satisfied by the construction of $d$.
\qed

Let $\calt$ be a (nondifferential) operad.
Its dual bar construction $\Omega(\calt)$ is of the form
$(\free(Z),d_\Omega)$ with $Z := \#\desusp \calt$ and $d_\Omega =
d_E$,
the internal part $d_I$ of the differential $d_\Omega$ being zero as
$\calt$ has trivial differential. The collection
$Z$ decomposes as $Z =
Z^0\op Z^1 \op \cdots$, where the collections
$Z^k$ are, for $k\geq 0$, defined by
\[
Z^k(n):=
\left\{
\begin{array}{ll}
Z(n),&\mbox{for } n=k+2, \mbox{ and}
\\
Z(n)=0,& \mbox{otherwise.}
\end{array}
\right.
\]
Then clearly $d_\Omega(Z^k)\subset \free^{k-1}(Z)$. Let $\cals$ be the
operad defined by
\[
\cals := \free(Z^0)/\Im(d_\Omega :\free^1(Z) \to \free(Z^0))
\]
and let $\rho : \Omega(\calt) \to \cals$ be the obvious map.
The following proposition is more or less the definition of the
Koszulness as it
is given in~\cite{GK}.

\begin{proposition}
\label{oslicek}
The object
\[
\rho : \Omega(\calt) \to \cals
\]
constructed above is a bigraded model for $\cals$
if and only if the operad
$\cals$ is Koszul. In this case $\pi(\cals) =
\calt = \ssusp\cals^!$, where $\ssusp\cals^!$ denotes the suspension
of the Koszul dual of the
operad $\cals$ (see~{\rm\cite{GK}} for the
definition of the Koszul dual).
\end{proposition}

The proposition above explicitly describes the structure of bigraded
models for Koszul operads. A nice thing is that
the most important examples of operads encountered in life
are Koszul, in contrast with
topology, where the ``Koszul spaces''
(i.e.~the spaces such that the universal
enveloping algebra of the homotopy Lie algebra is Koszul) are not of
much
interest. The following, though a bit artificial, example shows that
some
explicit computation is possible also for non-Koszul operads.

\begin{example}{\rm\
\label{fuk}
All the objects considered in this example are {\em nonsymmetric\/}.
Let
$\cals$ be the operad defined as $\cals := \free(E)/(I)$
with $E = E(3) =\Span(y) $ and the ideal $I$ generated by $y\comp_1
y\in \free(E)(5)$. Define the collection $Z = Z^0\op Z^1\op \cdots$ by
\[
Z^k(n):=
\left\{
\begin{array}{ll}
\Span(\eta_k),&\mbox{for } n=2k+3, \mbox{ and}
\\
0,& \mbox{otherwise,}
\end{array}
\right.
\]
where $\eta_k$ are generators of inner degree $k$, $k\geq 0$,
let the differential $d$ on $\free(Z)$ be given by
\[
d(\eta_k):=\sum_{a+b=k-1}(-1)^{a}
\eta_a\comp_1\eta_b,
\]
and the map $\rho:(\free(Z),d)\to
\cals$ by $\rho(\eta_0):=y$ and $\rho(\eta_k):=0$ for $k\geq 1$. A
direct
computation shows that $\rho:(\free(Z),d)\to
\cals$ is a bigraded model of $\cals$.

The operad \S\ is manifestly {\em not} Koszul, since the definition of
the Koszulness as it is given in~\cite{GK,GJ} requires the
quadraticity, which is not the case of \S.

The associated homotopy
operad $\calt$ can be described as follows.
Let
\[
\calt(n):=
\left\{
\begin{array}{ll}
\Span(\xi_k),&\mbox{for } n=2k+3, \mbox{ and}
\\
0,& \mbox{otherwise,}
\end{array}
\right.
\]
where $\xi_k$ are generators of inner degree $k+1$, $k\geq 0$,
and define the structure maps by
\[
\xi_{k}\comp_i\xi_{l}:=
\left\{
\begin{array}{ll}
\xi_{(k+l)},&\mbox{for } i=1, \mbox{ and}
\\
0,& \mbox{otherwise}
\end{array}
\right.
\]
The operad $\calt$ can be presented also as $\calt := \free(F)/(J)$,
where $F = F(3) = \Span(z)$ is spanned by a generator $z$ of inner
degree $1$ and the ideal $J$ is generated by
$z\comp_2 z, z\comp_3z \in \free(F)(5)$.
}\end{example}

Let us prove the following technical lemma (an analog
of~\cite[Lemma~4.5]{HS}).

\begin{lemma}
\label{Bucuresti}
Let $\rho :(\free(Z),d)\to(\calh,0)$ be a TJ model of an operad
$\calh$ as in Theorem~\ref{TJ}. Suppose that $\eta:
\calh \to \free(Z^0)$ is such that $\rho\eta =
\id$. Suppose that $D$ is a differential
on $\freez{\leq n}$ such that
$(D-d) : Z^l \to \free(Z)^{\leq l-2}$ for each $0\leq l \leq n$.
Then there exist equivariant
maps $v_n :Z(\free(Z)^{\leq n-1},D)
\to \free(Z)^{\leq n}$ and $a_n:Z(\free(Z)^{\leq n-1},D) \to \calh$
such
that
\[
u= Dv_n(u)+\eta(a_n(u)),\mbox{ for any $u\in Z(\free(Z)^{\leq
n-1},D)$}.
\]
\end{lemma}

\noindent{\bf Proof.}
For any $u \in \freez0 = Z(\free(Z)^{\leq 0},D)$ define
$a_1(u):= \cl(u)$. Because $\cl(u) = \cl(\eta a_1(u))$,
there obviously exists
a map $v_1 :\freez0 \to \free(Z)^{\leq 1}$ such that
\[
u-\eta(a_1(u))=d v_1(u) = D v_1(u)
\]
and we may suppose, using the same trick of the symmetrization
as in~(\ref{symmetrization}),
that the map $v_1$ is equivariant.

Suppose we have already proved our theorem for $n-1$, $n\geq 2$. Any
$D$-cycle $u\in Z(\free(Z)^{\leq n-1},D)$ decomposes as
$u=\sum_{j=0}^{n-1}u_j$, $u_j \in \free(Z)^j$. Then obviously
$du_{n-1} = 0$ and we may find some $w_n\in \free(Z)^n$ with
$u_{n-1}=dw_n$.
Now apply the induction assumption on the $D$-cycle $(u-Dw_n)\in
\free(Z)^{\leq n-2}$ to see that
\[
u-Dw_n = Dv_{n-1}(u-Dw_n)+\eta(a_{n-1}(u-Dw_n)).
\]
We see (using the same trick as in~(\ref{symmetrization}))
that $w_n$ can be chosen to depend equivariantly on $u$,
and we may define
$v_n(u): = w_n(u)+v_{n-1}(u-Dw_n(u))$ and $a_n:= a_{n-1}(u-Dw_n(u))$.
\qed

In the following theorem we show that for a differential operad there
exists an analog of the filtered model of~\cite[par.~4]{HS}.

\begin{theorem}
\label{filtered}
Let $(\cals,d_\cals)$ be a differential operad (with $\cals(1)=0$)
and let $\calh := \calh(\cals,d_\cals)$
be its homology. Let $\rho :(\free(Z),d)\to(\calh,0)$ be the
TJ-model of $\calh$ as in Theorem~\ref{TJ}.
Then there exists a differential $D$
on $\free(Z)$ and a homomorphism
$\alpha :(\free(Z),D)\to (\cals, d_\cals)$
such that
\begin{enumerate}
\item[(i)]
the differential $D$ is a perturbation of the
differential $d$ in the sense that
$(D-d)(Z^k)\subset \free(Z)^{\leq k-2}$ for any $k\geq 0$, and
\item[(ii)]
the map $\alpha$ induces an isomorphism
$\calh(\alpha):\calh(\free(Z),D)\to
\calh(\cals, d_\cals)$.
\end{enumerate}
An object $\alpha :(\free(Z),D)\to (\cals, d_\cals)$ as above is
called the
filtered model of the differential operad $(\cals,d)$.
\end{theorem}

\noindent{\bf Proof.}
Fix an equivariant map $\eta :\calh \to \freez0$ with $\rho\eta =
\id$.
Let $D:=0$ on $\freez0$ and let $\alpha|_{Z^0}$ be defined by the
commutativity of the diagram
\[
\Htriangle%
{Z^0}{Z(\cals,d_\cals)}{\calh}%
{\alpha|_{Z^0}}{\cl}{\rho}
\]
Put $D:= d$ on $Z^1$. Then $\cl(\alpha D)=\cl (\alpha d)=0$ on $Z^1$
and we can define
$\alpha|_{Z^1}$ by the commutativity of the diagram
\[
\square%
{Z^1}{\cals}{\freez0}{\cals}{\alpha|_{Z^1}}D{d_\cals}{\alpha}
\]

Suppose that $z\in Z^2$, then $dz \in \free^1(Z)$, hence $Ddz = d^2z
= 0$ and $d_\cals(\alpha dz)=0$. Extend $D$ by $D(z): = d(z)
-\eta(\cl(\alpha dz))$,
for $z\in Z^2$. Since, for any $a\in \calh$, $\cl(\alpha(\eta(a))) =
\rho\eta(a) = a$,
\[
\cl(\alpha Dz)= \cl(\alpha dz)- \cl(\alpha\eta(\cl(\alpha dz))) =
0,
\]
for all $z\in Z^2$. This means that we can extend $\alpha$ by the
commutativity of the diagram
\[
\square%
{Z^2}{\cals}{\free(Z)^{<2}}{\cals}{\alpha|_{Z^2}}D{d_\cals}{\alpha}
\]

Suppose we have already extended $D$ and $\alpha$ to $\free(Z)^{\leq
n}$ for some $n\geq 2$. For any $u\in Z^{n+1}$, $D(du)$ is a $D$-cycle
in $\free(Z)^{\leq n-1}$ and Lemma~\ref{Bucuresti} gives
\[
D(du) = Dv_{n}(u)+\eta(a_{n}(u)).
\]
Applying $\alpha$ to the above equation we get
\[
d_\cals \alpha(du) = d_\cals\alpha(v_{n}(u))
+\alpha\eta(a_{n}(u)),
\]
which implies that $\cl(\alpha\eta(a_{n}(u))= a_{n}(u)=0$,
therefore $D(du - v_{n}(u))=0$. Extend $D$ on $Z^{n+1}$ by
\[
Du = du - v_{n}(u)-\eta(\cl \alpha(du - v_{n}(u)).
\]
then obviously $\cl(\alpha Du)=0$ and we can extend $\alpha$ to
$Z^{n+1}$ so that the diagram
\[
\Square%
{Z^{n+1}}{\cals.}{\free(Z)^{<{n+1}}}%
{\cals}{\alpha|_{Z^{n+1}}}D{d_\cals}{\alpha}
\]
commutes.

To prove that $\calh(\alpha)
:\calh(\free(Z),D)\to \calh(\cals,d_\cals)$ is an
isomorphism, observe that $\calh(\eta):\calh(\cals,d_\cals)\to
\calh(\free(Z),D)$ is an epimorphism by
Lemma~\ref{Bucuresti} while obviously
$\calh(\alpha)\calh(\eta)=\id$ which implies that $\calh(\eta)$ is an
isomorphism and $\calh(\alpha)$ its inverse.
\qed

Let $\cals$ be a {\em nonsymmetric\/} operad
(possibly differential) and let
${\cal M}$ be some of the models of $\cals$
constructed above made in the
category of {\em nonsymmetric\/} operads.
We can consider $\cals$ equally well as a
{\em symmetric\/} operad $\cals_\Sigma$
with the trivial action of the symmetric group.
Let ${\cal M}_\Sigma$ be the
corresponding model for $\cals_\Sigma$ in the category of {\em
symmetric
operads\/}. We may then formulate the following principle.

\begin{odstavec}
{\rm Imbedding Principle.}
The models ${\cal M}$ and ${\cal M}_\Sigma$ are isomorphic, ${\cal
M}\cong
{\cal M}_\Sigma$.
\end{odstavec}

Loosely speaking, the above principle says that if
$\cals$ is an operad whose
axioms can be formulated without making use of the
symmetric group action (for
example the operad {\it Ass\/} for associative
algebras or the operads from
Example~\ref{fuk}), then all the ``higher syzygies''
of $\cals$ do not contain
the action of the symmetric group as well.

\section{Cotangent cohomology}

In~\cite{M} we constructed, for an arbitrary equationally given
category, a cohomology theory (called the {\em cotangent cohomology\/}
to stress the analogy with the cotangent cohomology of a
commutative algebra) as ``the best possible'' cohomology theory
controlling the deformations of the objects of this category. Our
cohomology was defined only in small, relevant degrees, and we raised
the question when this cohomology can be naturally extended in all
degrees. In this paragraph we show that such an extension exists for
an {\em algebraic\/} equationally given category, i.e.~for a category
of algebras over some operad. We indicate also some applications and
connections with~\cite{FM,GK}.

Let $\cals$ be an operad and $M$ an \S-module
(see Definition~\ref{module}).
By a {\em degree $p$ derivation\/} of the operad $\cals$ in $M$
we mean a sequence $\theta
=\{\theta(n):\cals(n)\to M(n);\ n\geq 2\}$ of equivariant degree $p$
linear maps such that
\[
\theta(m+n-1)(f\comp_i g) =
\theta(m)(f)\comp_i^R g + (-1)^{p\cdot|f|}\cdot f\comp_i^L
\theta(n)(g),
\]
for any $f\in \cals(m)$, $g\in \cals(n)$, $m,n\geq 1$ and $1\leq i\leq
n$.
We denote by $\der_p(\cals,M)$ the vector space of all degree $p$
derivations of $\cals$ in $M$.

Let $\rho:(\free(Z),d) \to (\cals,0)$ be a bigraded model of \S\ as in
Theorem~\ref{TJ}. If $M$ is an \S-module, then the homomorphism $\rho$
induces on $M$ a structure of an $\free(Z)$-module and we may consider
the space $C^*(\cals;M) := \Der_*(\free(Z),M)$ of derivations of
$\free(Z)$ in $M$.

\begin{proposition}
The restriction induces an isomorphism $C^*(\cals;M) \cong \mbox{\rm
Coll}_*(Z,M)$. If we denote
\[
\csm p*:= \{\theta \in C^*(\cals;M);\ \theta|_{Z^k}=0 \mbox{ for }
k\not= p\},
\]
then the formula $\nabla(\theta) := \theta \circ d$ gives a
well-defined endomorphism $\nabla$ of $C^*(\cals,M)$ with
\[
\nabla(\csm pq) \subset \csm{p+1}{q-1},
\]
and, moreover, $\nabla$ is a differential, $\nabla^2=0$.
\end{proposition}

\noindent{\bf Proof.}
The isomorphism is a consequence of the universal property of
$\free(Z)$. The only thing which has to be checked carefully is that
$\nabla(\theta)$ is a derivation. Let us verify this.
Let $\theta \in \csm pq$,
$f\in \free(Z)(m)$, $g\in \free(Z)(n)$ and $1\leq i\leq n$.
Then
\begin{eqnarray*}
&&\nabla(\theta)(f\circ_i g) = \theta d(f\circ_i g)
= \theta(df\circ_i g) +(-1)^{|f|}\cdot \theta (f\circ_i dg)
=\theta(df)\comp_i^R \rho(g)+
\\
&&\hphantom{mmmmm} + (-1)^{q(|f|+1)}\cdot \rho(df)\comp_i^L
\theta(g) + (-1)^{|f|}\cdot \theta(f)\comp_i^R \rho(dg)
+(-1)^{|f|(q+1)}\cdot\rho(f)\comp_i^L \theta(dg)
\\
&&\hphantom{mmmmm} =\theta(df)\comp_i^R
\rho(g)+ (-1)^{|f|(q+1)}\cdot\rho(f)\comp_i^L
\theta(dg)
\\
&&\hphantom{mmmmm} = \nabla(\theta)(f)\comp_i^R \rho(g)+
(-1)^{|f|(q+1)}\cdot\rho(f)\comp_i^L
\nabla(\theta)(g),
\end{eqnarray*}
because $\rho \circ d=0$.
The condition $\nabla^2=0$ is easy to verify,
$\nabla^2(\theta) = \nabla(\theta)(d)
= \theta(d^2) =0$.
\qed

\begin{definition}
\label{cotangent}
Let $\cals$ be an operad and $M$ an $\cals$-module. Then the cotangent
cohomology of $\cals$ with coefficients in $M$, $\cotsm**$, is defined
as
\[
\cotsm** := H^{*,*}(\csm**,\nabla).
\]
\end{definition}

The following two definitions are motivated by one of the most
important
concepts of rational homotopy theory, by the notion of the formality.
We say that a differential operad $(\calt,d_\calt)$ is {\em formal\/}
if it has the same minimal model as its homology operad
$\calh(\calt,d_\calt)$. A (nondifferential) operad $\calh$ is said to
be {\em intrinsically formal\/} if any differential operad
$(\calt,d_\calt)$ with $\calh(\calt,d_\calt)\cong \calh$ is formal.
The following proposition shows that intrinsic formality is
obstructed by the cotangent cohomology constructed above, compare the
corresponding statement of~\cite{HS}.

\begin{theorem}
\label{obstruction}
Any (nondifferential) operad $\calh$ with $\cot{\geq
2}{-1}(\calh,\calh)=0$ is intrinsically formal.
\end{theorem}

\noindent{\bf Proof.}
Let $\rho:(\free(Z),d) \to (\calh,0)$ be a bigraded model of $\calh$.
Consider the space $\Der_*({\free(Z)})$ of derivations of $\free(Z)$
with the differential $\Delta$ defined by $\Delta(\phi):= d\circ \phi
-
(-1)^{|\phi|}\phi\circ d$. We have on $\Der_*{\free(Z)}$ a second
grading
given by
\[
\Der^q_*(\free(Z)):=\{\phi \in \Der_*(\free(Z));\
\phi(Z^k)\subset \free^{k-q}(Z),\ k\geq 0\}
\]
and $\Delta(\Der^q_p(\free(Z))\subset \Der^{q+1}_{p-1}(\free(Z))$.
Define
$\chi : \Der^*_*(\free(Z))\to C^{*,*}(\calh,\calh)$ by
$\chi(\phi):=(-1)^{|\phi|}\cdot \rho\circ \phi$. We have
\[
\chi(\Delta(\phi)) = \chi( d\phi -
(-1)^{|\phi|}\phi d) = \chi(\phi)\circ d = \nabla(\chi(\phi)),
\]
so $\chi$ is a map of differential spaces, $\chi:
(\Der^*_*(\free(Z)),\Delta)\to (C^{*,*}(\calh,\calh),\nabla)$ and a
spectral sequence argument based on the filtration induced by
the TJ-grading shows that $\chi$
is a homology isomorphism. We may now reformulate the hypothesis of
the theorem as
\begin{equation}
\label{myska}
H^{\geq 2}_{-1}(\Der^*_*(\free(Z)),\Delta)=0.
\end{equation}
Let $(\calt,d_\calt)$ be an operad such that $\calh \cong
\calh(\calt,d_\calt)$. By Theorem~\ref{filtered} the
filtered model of $(\calt,d_\calt)$ is of the form $(\free(Z),D)$,
where $D$ is a perturbation of $d$ in the sense that $D =
d+d_2+d_3+\cdots$, $d_k\in \Der^k_{-1}(\free(Z))$. Similarly as
in~\cite{HS} we may see that the triviality of such a perturbation is
obstructed by (part of) the cohomology of~(\ref{myska}), therefore the
filtered
model of $(\calt,d_\calt)$ is isomorphic with the bigraded model of
$\calh$ which is minimal, by definition.
\qed

As a consequence of our computations we get the following stunning
result.

\begin{proposition}
\label{stunning}
Let $\calh$ be a (nondifferential) operad such that $\calh(n)$ is
concentrated in degree $N(n-1)$, for some $N$ and any $n\geq 2$. Then
$\calh$ is intrinsically formal.
\end{proposition}

\noindent{\bf Proof.}
Let $\rho : (\free(Z),d)\to \calh$ be the bigraded model of $\calh$.
We show that the collection $Z^k$ has the property that $Z^k(n)$ is
concentrated in degree $N(n-1)+k$, for any $n\geq 2$ and $k\geq 0$.
Let us prove this statement inductively.

For $k=0$, the collection $Z^0$ consists of generators of $\calh$,
therefore $Z^0(n)$ is concentrated in degree $N(n-1)$ by the
assumption. Suppose we have
proved the statement for any $k< l$, for some $l\geq 1$. We prove that
${\cal F}(Z^{<l})^{l-1}(n)$ is concentrated in degree $N(n-1) + (l-1)$
which would give the induction step, since $Z^l$ was constructed as
the suspension of the space of generators of
$\calh^{l-1}(\free(Z^{<l}),d)$, see the proof of Theorem~\ref{TJ}.

Any element $w\in {\cal F}(Z^{<l})^{l-1}(n)$ is a sum of elements of
the form
\[
w_1 \circ_{j_1}(w_2 \circ_{j_2} (\cdots \circ_{j_{p-1}} w_p)\cdots),
\]
for some $p\geq 2$, $w_i \in Z^{s_i}(n_i)$, with $s_1+\cdots+s_p =
l-1$ and $n_1+\cdots n_p = n+p-1$. By the induction assumption,
$\deg(w_i) = N(n_i-1)+s_i$, $1\leq i\leq p$, therefore
\[
\deg(w) = \sum_{i=1}^p (N(n_i-1)+s_i) = N(n-1)+(l-1)
\]
as claimed.

Suppose $\theta \in C^{\geq2,-1}(\calh,\calh)$. We show that $\theta =
0$. It is enough to verify it on generators, so let us pick some $z\in
Z^k(n)$; we already know that $\deg(z) = N(n-1) +k$. Since $\theta(z)
\in \free^{\leq k-2}(n)$, $\theta(z)$ must be a sum of elements of the
form
\[
w_1 \circ_{j_1}(w_2 \circ_{j_2} (\cdots \circ_{j_{p-1}} w_p)\cdots),
\]
for some $p\geq 2$, $w_i \in Z^{s_i}(n_i)$, with $s_1+\cdots+s_p \leq
k-2$ and $n_1+\cdots +n_p = n-p+1$. The similar degree argument as
above gives that $\deg(\theta(z)) \leq N(n-1) + k-2 = \deg(z)-2$,
which
is impossible, since we should have $\deg(\theta(z))=\deg(z)-1$.

We proved that $C^{\geq2,
-1}(\calh,\calh)=0$, therefore
the relevant part of the cohomology of Theorem~\ref{obstruction}
vanishes.
\qed

Notice that an equivalent way to formulate the hypothesis of
Proposition~\ref{stunning} is to say that the operad $\calh$ is an
$N$-fold suspension $\ssusp^N \calr$ of a nongraded (= concentrated in
degree zero) operad~$\calr$.

\begin{example}{\rm\
In this example we construct a differential operad which is not
formal. Let $E = E(2)\op E(4)$ be the collection given by $E(2):=
\Span(f)$, $E(4):= \Span(g)$, where $f$ (resp.~$g$) is a generator of
inner degree~$1$ (resp.~$4$). Let $\calt := \free(E)$ and define a
differential $d_\calt$ on $\calt$ by
\[
d_\calt(f):= 0\mbox{ and }
d_\calt(g) := f\comp_1 f\comp_1 f.
\]
Then the differential operad $\Omega(\calt,d_\calt)$ is not formal.

To see this, recall that $\Omega(\calt,d_\calt)$ is of the form
$(\free(Z),d_\Omega)$ with $Z = \#\desusp \free(E)$. Let $Z^k :=
\#\desusp \free^{k+1}(E)\subset \#\desusp \free(E)= Z$. Then
$d_E(Z^k) \subset \free(Z)^{\leq k-1}$ (where
$d_E$ is the external part of the differential $d_\Omega$) and it is
not difficult to see that
$(\free(Z),d_E)$ is the bigraded model of the operad $\caly :=
\free(Z)/\free^{\geq 1}(Z)$, which is nothing else but the collection
$\desusp \# E$ with trivial structure maps. The differential $d_\Omega
= d_E + d_I$ is then a nontrivial perturbation of the ``formal''
differential $d_E$.

Therefore $\Omega(\calt,d_\calt)$ is a differential operad which is
not
formal, its homology operad is isomorphic to the operad $\caly$
described
above and its homotopy operad $\pi\Omega(\calt,d_\calt)$ is the
homology
operad $\calh(\calt,d_\calt)$.

A topological analog of this types of examples would be a nonformal
space obtained from a wedge of spheres by attaching a cell by a
suitable map.
}\end{example}

Let $A$ be an $\cals$-algebra, i.e.~a vector space $V$ together with a
homomorphism $A:\cals \to \End(V)$ and let $\bk_V$ be the operad
$\End(V)$ together with the $\cals$-module structure induced by $A$
(see Example~\ref{modules}).

\begin{definition}
\label{pik}
Define the \S-cohomology of $A$
(with coefficients in itself) as
\[
H^{*,*}_\cals(A) := \cot **(\cals;\bk_V).
\]
\end{definition}

It is possible to show that, for a {\em Koszul\/} operad $\cals$, the
cochain complex $(C^{*,*}(\cals),\nabla)$ used in the definition of
the above cohomology coincides with the cochain complex used in the
definition of the cohomology introduced in~\cite{GK} (or, more
precisely, with the dual form of this complex introduced
in~\cite{FM}). Thus the above cohomology is a natural extension
of the construction of~\cite{GK} for algebras over arbitrary, not
necessary Koszul, operads, hence an extension of the ``classical
cohomology'' as well.

Using a trick similar to the one in~\cite{FM} we may extend the
cohomology
of Definition~\ref{pik} for an arbitrary coefficient module $N$
(denote this extension by $H^{*.*}_\cals(A;N)$) and
we may show that $H^{*.*}_\cals(A;N)$ is trivial
whenever $A$ is a free $\cals$-algebra. A theorem of~\cite{B} then
says that $H^{*.*}_\cals(A;N)$ coincides with the Barr-Becker
cohomology of
$A$ with coefficients in $N$ (notice that the category of
$\cals$-algebras is obviously tripleable).

The above observation may seem to be an extremely strong result. For
example, let \comm\ be the operad for the category of commutative
algebras. This operad is
Koszul (see~\cite{GK}), hence the cohomology of Definition~\ref{pik}
coincides with the cohomology of~\cite{GK} which is in this case
manifestly isomorphic to the Harrison cohomology. Our claim
about the vanishing of $H^{*.*}_\cals(A;N)$ implies that the Harrison
cohomology of the polynomial algebra over a field of characteristic
zero is trivial, for arbitrary coefficients.
This is a very difficult result and we do not know any
reasonably simple proof of this statement, which was proved by M.~Barr
using the Hodge decomposition trick. The point is that the Koszulness
of \comm\
is in fact equivalent with the vanishing property for the Harrison
cohomology, as it was pointed out in~\cite{GK}.

\section{Homotopy $\cals$-algebras, homotopy everything spaces}

By a {\em homotopy algebra\/} we mean an algebra over a {\em
minimal\/} (in the sense of Theorem~\ref{minimal}) differential
operad. If $A : (\free(M),d)\to (\End(V),\dend)$ is a homotopy algebra
structure on $(V,d_V)$ and $(\free(M),d)$ is a minimal model of some
$(\cals,d_\cals)$ we sometimes say also that $A$ is a {\em homotopy\/}
$(\cals,d_\cals)$-{\em algebra\/} structure on $(V,d_V)$.

\begin{example}{\rm\
\label{ferda}
If the operad $\cals$ is Koszul, then $(\free(Z),d) =
\Omega(\ssusp\cals)$
by~Proposition~\ref{oslicek} and our definition coincides with that
of~\cite{GK}. Especially, homotopy associative algebras are
A($\infty$)-algebras of~\cite{St}, homotopy commutative associative
algebras are
the balanced A($\infty$)-algebras of~\cite{K,Ma} while
homotopy Lie algebras are
strong Lie homotopy algebras of~\cite{HS1,LS,LM}.
}\end{example}

\begin{example}{\rm\
Let us consider the nonsymmetric operad
$\cals$ of Example~\ref{fuk}. An $\cals$-algebra on a
differential graded vector
space $(V,d_V)$ is given by a degree zero
map $\mu : V^{\ot3}\to V$ which commutes with
the differential and satisfy $\mu(\mu\ot\id^2)=0$.

A homotopy $\cals$-algebra is then a differential space $(V,d_V)$
together with degree $k$ linear maps $h_k :V^{\ot(2k+3)}\to V$, $k\geq
0$,
satisfying
\begin{eqnarray*}
0&=& dh_0 + h_0(d\ot \id^2+\id\ot d\ot\id + \id^2\ot d),
\\
h_0(h_0\ot \id^2)
&=& dh_1 - h_1(d\ot \id^4+\cdots + \id^4\ot d),
\\
-h_1(h_0\ot \id^4)+h_0(h_1\ot \id^2)&=&
dh_2 + h_2(d\ot \id^6+\cdots + \id^6\ot d),
\\
&:&
\\
\sum_{i+j=k}(-1)^j \cdot h_i(h_j\ot \id^{(2i+2)})
&=& dh_k +(-1)^k\cdot h_k(d\ot \id^{(2k+2)}+\cdots + \id^{(2k+2)}\ot
d).
\end{eqnarray*}
}\end{example}

\begin{definition}
Let $(\cals,d_\cals)$ be a differential operad, $A:(\cals,d_\cals) \to
(\End(V),\dend)$ and $(\cals,d_\cals)$-algebra structure on a
differential space $(V,d_V)$ and $\nu : (\free(M),d)\to
(\cals,d_\cals)$ the minimal model of $(\cals,d_\cals)$. We then call
the structure $A\circ \nu :(\free(M),d) \to (\End(V),\dend)$ the
homotopy structure associated to $A$.
\end{definition}

Let us discuss the possibility of an intrinsic characterization of the
associated homotopy structure defined above. We first state the
following very natural definition.

Let $\calt$ and $\calu$ be two (nondifferential)
operads, let $H$ be a graded vector space and let $a: \calt \to
\End(H)$ and $b: \calu \to
\End(H)$ be two structures on $H$. We say that the structures $a$ and
$b$ are {\em equivalent\/} (= are the same) if there exists an operad
isomorphism $\varphi : \calt\to \calu$ such that the diagram
\[
\Ztriangle{\calt}{\End(H)}{\calu}{\varphi}{b}{a}
\]
commutes. The \underline{first} important
property of the associated homotopy
structure is that it induces on $H = H(V,d_V)$ the same homology
structure as the operad $(\cals,d_\cals)$. This is immediately
obvious from the diagram
\[
\Ztriangle{\calh(F(M),d)}%
{\hphantom{xxxxxxxxxxxxxxxxxxx}\calh(\End(V),\dend)\cong(\End(H),0)}%
{\calh(\cals,d_\cals)}%
{\calh(\nu)}{\calh(A)}{\calh(A\circ \nu)}
\]
In this context, it is interesting to formulate the following
observation.

\begin{observation}
Let $A:(\cals,d_\cals)\to (\End(V),\dend)$ be an
$(\cals,d_\cals)$-algebra structure on $(V,d_V)$ and denote $\calh :=
\calh(\cals,d_\cals)$. Then there exists on $(V,d_V)$ an
$\calh$-algebra structure inducing on $(V,d_V)$ the same homology
structure as $A$.
\end{observation}

To see this, let $\Phi : (\End(H),0) \to (\End(V),\dend)$ be the
(noncanonical) homomorphism constructed in Lemma~\ref{micinka}. Then
$\Phi
\circ \calh(A)$ is the requisite $\calh$-algebra structure.

The \underline{second} important property of the associated homotopy
algebra structure is that it factors through $A$. This means, loosely
speaking, that both the operations and the axioms of the associated
homotopy structure are expressible in terms of compositions of the
operations of the $(\cals,d_\cals)$-structure. Using the terminology
borrowed from the universal algebra we may say that the associated
homotopy algebra structure is a {\em specialization\/} of $A$.

The \underline{third} important property of the associated homotopy
algebra structure is the following universality. Let $(\calu,d_\calu)$
is another differential operad which satisfies the above conditions,
then there exists a homomorphism $\psi :(\free(M),d) \to
(\calu,d_\calu)$ inducing the isomorphism in homology such that the
diagram
\[
\Ztriangle{(\free(M),d)}{(\cals,d_\cals)}%
{(\calu,d_\calu)}{\psi}{}{\nu}
\]
commutes up to homotopy. Of course, strictly speaking, we have no
right to formulate this statement as we never have introduced the
notion of a homotopy for maps of operads (see also the comments in the
introduction), but we may at least claim that the three properties
listed above determines uniquely the associated homotopy
algebra structure up
to an acyclic factor.

The following proposition shows the relevance of the notion of the
(intrinsic) formality introduced in the previous paragraph.

\begin{proposition}
\label{kozulka}
Suppose that the differential operad $(\cals,d_\cals)$ is formal and
let $\calh = \calh(\cals,d_\cals)$ be its homology operad. Then
every $(\cals,d_\cals)$-algebra admits an associated
homotopy $\calh$-algebra structure.

The above statement is in particular true for any operad
$(\cals,d_\cals)$ whose homology operad $\calh =
\calh(\cals,d_\cals)$ is concentrated in degree zero or which is an
(iterated) suspension of such an operad.
\end{proposition}

\noindent{\bf Proof.}
Let $A: (\cals,d_\cals)\to (\End(V),\dend)$ be an $ (\cals,d_\cals)$
algebra structure on $(V,d_V)$. By the definition of
formality, the minimal model of $(\cals,d_\cals)$ is the same as the
minimal model (= the bigraded model) $\nu:(\free(M),d)\to (\calh,0)$
of $\calh$. Let $\alpha : (\free(M),d)\to (\cals,d_\cals)$ be the
corresponding minimal model map.
The composition $A\circ \alpha : (\free(M),d)\to
(\End(V),\dend)$ is then the desired associated $\calh$-homotopy
algebra structure.
The second part follows from Proposition~\ref{stunning}.
\qed

Having in
mind future applications we generalize our definitions a bit.
Let $(\cals,d_\cals)$ be a differential operad with the condition
$\cals(1)=0$ replaced by $H(\cals(1),d_\cals(1))=0$. Let
$(\st,d_\st)$ be its universal covering constructed in section~1. The
inclusion $\iota: (\st,d_\st) \hookrightarrow (\cals,d_\cals)$ is a
differential operad map inducing an isomorphism in homology and for
any $(\cals,d_\cals)$-algebra
structure there exists a natural $(\st,d_\st)$-algebra
structure induced from the former one by $\iota$.
In the rest of the paper, by an associated homotopy algebra
structure we mean a homotopy structure associated to this induced
$(\st,d_\st)$-algebra structure.

\begin{example}{\rm\
Let $(\calo,d_\calo)$ be an operad such that $\calh(\calo,d_\calo)$ is
isomorphic to the operad \comm\ for commutative algebras. Then any
$(\calo,d_\calo)$-algebra structure admits an associated
balanced A($\infty$)-algebra structure, see~\cite{Ma} for the
definition.
This is in particular true for so-called
May algebras (see~\cite{HS2}) which are, by definition, acyclic
operads augmented over the operad \comm.
Our statement follows from the fact that the
operad \comm\ is concentrated in
degree zero (notice that \comm\ is, in a sense, a trivial symmetric
operad, $\comm(n) = \bk$ for any $n \geq 2$)
and from the fact that homotopy \comm-algebras are balanced
A($\infty$)-algebras, see the discussion in Example~\ref{ferda}.
A nonsymmetric analog of the above claim is the following.

Let $(\calu,d_\calu)$ be a nonsymmetric
operad such that $\calh(\calu,d_\calu)\cong \ass$, where \ass\ denotes
the nonsymmetric operad for the category of associative algebras.
As above,
the operad \ass\ is, in a sense, the trivial nonsymmetric operad,
$\ass(n) = \bk$ for any $n\geq 2$.
Any $(\calu,d_\calu)$-algebra then admits an associated
A($\infty$)-algebra structure. This again follows from the fact that
the operad \ass\ is nongraded and from the remarks of
Example~\ref{ferda}.
}\end{example}

\begin{example}{\rm\
In this example we show that both the bigraded model of an operad and
the associated homotopy algebra structure is implicitly hidden
in the 30 years old papers~\cite{St}.
Recall that the {\em associahedron\/} $K_n$ is, for
$n\geq 2$, an $(n-2)$-dimensional
polyhedron whose $i$-dimensional cells are, for $0\leq i\leq n-2$,
indexed by all (meaningful) insertions of $(n-i-2)$ pairs of brackets
between $n$ independent indeterminates, with suitably defined
incidence maps. There exists a natural
structure of a nonsymmetric topological operad on the nonsymmetric
collection $\calk := \{ K_n;\ n\geq 2\}$. A topological space $X$ is
called an A($\infty$)-{\em space\/} if it admits an action of
$\calk$, see~\cite{St} for details.

On the other hand, let \ass\ be the nonsymmetric operad for
associative algebras, i.e.~the operad with $\ass(n)=\bk$ for any
$n\geq 2$ and all the structure operations equal to the
identity, $\comp_i =
\id :\bk \ot \bk \cong \bk \to \bk$. Let $(\free(Z),d)\to (\ass,0)$ be
its bigraded model. We know from Proposition~\ref{oslicek}, and from
the
fact that the operad \ass\ is Koszul with $\ass^! = \ass$
(see~\cite{GK}) that
\[
(\free(Z),d)\cong (\Omega(\ssusp \ass),d_\Omega),
\]
and we can obtain from this the following very explicit description of
the bigraded model $(\free(Z),d)$:
there are some $\xi_k \in Z(k+2)$ of inner degree $k$ such that
$Z^k = \Span(\xi_k)$, $k\geq 0$. The differential $d$ is given by the
formula
\[
d(\xi_k) = \sum (-1)^{(a+1)(i+1)+a}\cdot \xi_a \comp_i \xi_b,
\]
where the summation is taken over all $a,b\geq 0$ with $a+b=k-1$ and
$1\leq i\leq b+2$. Due to the integrality of the coefficients at the
right-hand side of the above equation, we can consider an
obvious integral variant of the above construction, $(\fint(Z),d)$.

Let $(CC(\calk),d_C)$ be the nonsymmetric operad of
cellular chains on $\calk$,
i.e.~the operad given by $(CC(\calk),d_C)(n):= (CC(K_n),d_C)$ and the
structure maps induced by those of $\calk$; here $(CC(-),d_C)$ denotes
the cellular chain complex functor. According to the definition of the
associahedra recalled above, there is, for any $n\geq 2$, exactly one
top-dimensional cell $u_n$ in $K_n$ (given by the insertion of no pair
of
brackets between $n$-indeterminates and having
dimension = $(n-2)$) and it is a very stimulating exercise to prove
that the map $J: (\fint(Z),d)\to (CC(\calk),d_C)$ given by
\[
\fint(Z)\ni \xi_n \longmapsto 1\cdot u_n \in CC(\calk)
\]
is in fact an {\em isomorphism\/} of differential operads.

Suppose that a topological space $X$ admits an action of $\calk$. Then
there exists an induced action of the singular chain operad
$(CS(\calk),d_S)$ on the singular chain complex $(CS(X),d_S)$
of $X$ and the inclusion
$(CC(\calk),d_C)\hookrightarrow (CS(\calk),d_S)$ induces on
$(CS(X),d_S)$ a structure of an $(CC(\calk),d_C)$-algebra. This is
exactly the A($\infty$)-structure whose existence is proved
in~\cite{St}. Summing up the above remarks we see that
this structure is in fact an integral variant of
the associated homotopy algebra structure discussed in this paper.

We may say also that the existence of the associated
A($\infty$)-structure
was in~\cite{St} proven as a consequence of a very rich geometrical
structure on the corresponding topological operad.
Proposition~\ref{kozulka} says that structures of this type exist,
under fairly general assumptions, regardless of the concrete geometric
structure of the corresponding topological operad.

The above example also explains a close relationship between the
coherence implying the asphericity of the associahedra on one hand,
and the Koszulness of the operad \ass\ on the other hand,
the fact that puzzled us a lot in~\cite{M}.
}\end{example}

\begin{example}{\rm\
Let $\caln$ be the symmetric discrete topological operad with
$\caln(n)=P$ for $n\geq 2$, where $P$ is the one-point discrete
topological space. Any topological operad $\cale$ such that any
$\cale(n)$ is connected, $n\geq 2$, is obviously augmented over
$\caln$.

Suppose moreover that $\cale$ is acyclic (meaning that
$\cale(n)$ is acyclic for any $n\geq 2$). Then the singular chain
operad $(CS(\cale),d_S)$ is obviously augmented over $(CS(\caln),d_S)
= (\comm,0)$ and the augmentation induces a homology isomorphism.
If $X$ is a topological space on which such an operad acts,
then there exists an associated balanced A($\infty$)-algebra structure
on the singular chain complex $(CS(X),d_S)$ of $X$. This is in
particular
true for homotopy-everything spaces introduced in~\cite{May}.

There is an obvious nonsymmetric analog of the previous example.
Let $\calm$ be the nonsymmetric discrete topological operad with
$\calm(n)=P$ for $n\geq 2$. Any connected nonsymmetric topological
operad is clearly augmented over $\calm$.

If $\cala$ is a nonsymmetric aspheric topological operad, then the
singular nonsymmetric operad $(CS(\cala),d_S)$ is obviously augmented
over $(CS(\calm),d_S)
= (\ass,0)$ and the augmentation induces an isomorphism in
homology. If $Y$ is a topological space endowed with an action of
$\cala$, then the singular chain complex $(CS(Y),d_S)$ of $Y$ has an
associated A($\infty$)-algebra structure.
}\end{example}

\begin{example}{\rm\
\label{positron}
By an $(m,n)$-{\em algebra\/} we mean a (graded) vector space
$V$ together with two bilinear maps, $-\cup-:V \ot V \to
V$ of degree $m$, and $[-,-]:V\ot V\to V$ of degree $n$
($m$ and $n$ are natural numbers), such that,
for any homogeneous $a,b,c\in V$,
\begin{enumerate}
\item[(i)]
$a\cup b = (-1)^{|a|\cdot |b|+m}\cdot b\cup a$,
\item[(ii)]
$[a,b] = -(-1)^{|a|\cdot |b|+n}\cdot [b,a]$,
\item[(iii)]
$-\cup-$ is associative in the sense that
\[
a\cup (b\cup c) = (-1)^{m\cdot(|a|+1)}\cdot(a\cup b)\cup c,
\]
\item[(iv)]
$[-,-]$ satisfies the following form of the Jacobi identity:
\[
(-1)^{|a|\cdot(|c|+n)}\cdot [a,[b,c]]+(-1)^{|b|\cdot(|a|+n)}\cdot
[b,[c,a]]
+(-1)^{|c|\cdot(|b|+n)}\cdot [c,[a,b]] = 0,
\]
\item[(v)]
the operations $-\cup-$ and $[-,-]$ are compatible in the sense that
\[
(-1)^{m\cdot|a|}[a,b\cup c] =
[a,b]\cup c + (-1)^{(|b|\cdot |c|+m)}[a,c]\cup b.
\]
\end{enumerate}

Obviously $(0,0)$-algebras are exactly (graded) Poisson algebras,
$(0,1)$-algebras are Gerstenhaber algebras introduced in~\cite{GS}
while $(0,n-1)$-algebras
are the $n$-algebras of~\cite{GJ}. We may think of an
$(m,n)$-structure on
$V$ as
of a Lie algebra structure on $\uparrow^n V$
together with an associative commutative algebra
structure on $\uparrow^m V$ such that both structures are related via
the
compatibility axiom~(v).

Denote by $\calp(m,n)$ the operad for $(m,n)$-algebras, see \cite{FM}
for a very explicit description of this operad. In~\cite{FM}
we also computed the Koszul dual as $\calp(m,n)^! =\calp(n,m)$, it can
also
be proven, using the fact that $\calp(m,n)$ is an operad with a
distributive law and a suitable spectral sequence argument, that
$\calp(m,n)$ is Koszul (for the special case of $n$-algebras it was
done in~\cite{GJ}). Let $\cN$ be, for $N\geq 2$,
the ``$N$-little cubes operad''
of~\cite{BV} and let $(CS(\cN),d_S)$ be the
corresponding differential operad of singular chains. It is
known~\cite{GJ} that $\calh(CS(\cN),d_S) =\calp(0,N-1)$.

We do not know whether the operad $(CS(\cN),d_S)$ is formal (this
would imply that the singular chain complex $(CS(X),d_S)$ would have
an
associated homotopy $\calp(0,N-1)$-structure for any topological space
$X$ on which the little $N$-cubes operad $\cN$ acts). We show at least
that
$(CS(\cN),d_S)$ contains a formal suboperad with the homology
operad isomorphic to
$\ssusp^N \mbox{\it Lie}$, the $N$-fold suspension of the operad {\it
Lie\/} for Lie algebras.

Let $(\calr,d_\calr)$ be the differential suboperad of $(CS(\cN),d_S)$
defined by
\[
\calr_i(n): =
\left\{
\begin{array}{ll}
0,& \mbox{for $i< N(n-1)$,}
\\
\mbox{Ker}(d_S : CS_{N(n-1)}(\cN(n))\to CS_{N(n-1)-1}(\cN(n))),&
\mbox{for $i = N(n-1)$, and}
\\
CS_i(\cN(n)),& \mbox{for $i> N(n-1)$.}
\end{array}
\right.
\]
It is easy to see that $\calh(\calr,d_\calr) = \ssusp^N \mbox{\it
Lie}$
and we may apply Proposition~\ref{stunning} to
infer that the operad $(\calr,d_\calr)$ is formal. This means, by
definition, that the minimal model of $(\calr,d_\calr)$ is the same as
the minimal model of $(\ssusp^N \mbox{\it Lie},0)$ (which is
$\Omega(\ssusp^N \mbox{\it Comm}$)).

Let $X$ be a topological space on which the little $N$-cubes operad
$\cN$ acts. We have the induced action of $(CS(\cN),d_S)$ on
$(CS(X),d_S)$ and, via the inclusion $(\calr,d_\calr) \hookrightarrow
(CS(\cN),d_S)$, also the action of $(\calr,d_\calr)$ on
$(CS(X),d_S)$. The above shows that there is an associated homotopy
$\ssusp^N\mbox{\it Lie}$-action on the singular chain complex
$(CS(X),d_S)$ of $X$.
}\end{example}

In the following proposition we generalize a bit the arguments used in
the previous example.

\begin{proposition}
\label{trick}
Let $(\cals,d_\cals)$ be a differential operad and let $\calh =
\calh(\cals,d_\cals)$ be its homology operad. Suppose that there
exists a natural number $N$ such that the graded vector space
$\calh(n)$ is trivial in degrees $> N(n-1)$. Let $\cale$ be the
suboperad of the operad $\calh$ defined by $\cale_*(n) =
\cale_{N(n-1)}(n):= \calh_{N(n-1)}(n)$, for $n\geq 2$.

Then every $(\cals,d_\cals)$-algebra admits an associated homotopy
$\cale$-algebra structure.
\end{proposition}

\noindent
{\bf Proof.}
Let $(\calr,d_\calr)$ be the differential suboperad of
$(\cals,d_\cals)$ defined by
\[
\calr_i(n): =
\left\{
\begin{array}{ll}
0,& \mbox{for $i< N(n-1)$,}
\\
\mbox{Ker}(d_\cals : \cals_{N(n-1)}(n)\to \cals_{N(n-1)-1}(n),&
\mbox{for $i = N(n-1)$, and}
\\
\cals_i,& \mbox{for $i> N(n-1)$.}
\end{array}
\right.
\]
Then $\calh(\calr,d_\calr) = \cale$
and Proposition~\ref{stunning} implies the formality of the operad
$(\calr,d_\calr)$.
\qed

In the last two examples we discuss some applications of our methods
to
string theory. The terminology and the
notation (which may in some places collide with the notation we have
used so far) was taken from~\cite{KSV}; we refer to this paper for
more
details.

\begin{example}{\rm\
Let $\calp = \{\calp(n);\ n\geq 1\}$ be the collection defined
by $\calp(n)$ := the moduli space of nondegenerate Riemann
spheres with $(n+1)$ punctures and holomorphic disks at each
puncture. There exists an operad structure on $\calp$
induced by sewing Riemann spheres at punctures. Let us
recall that a (tree level) {\em conformal field theory\/} (CFT) based
on a state space $\calh$ is an action of the operad $\calp$ on the
topological vector space $\calh$, see~\cite[4.1]{KSV} for details.

Let us recall also that a {\em string background\/} is a CFT based on
a vector space $\calh$ with the following additional data.
\begin{enumerate}
\item[(i)]
A grading $\calh = \bigoplus \calh_i$ on the state space.
\item[(ii)]
An action of the Clifford algebra $C(V\oplus V^*)$ on $\calh$; here
$V$ denotes the complexification of the Virasoro algebra.
\item[(iii)]
A differential $Q:\calh\to \calh$ of degree $1$.
\end{enumerate}
These data are supposed to satisfy some axioms, see~\cite[4.2]{KSV}.
The differential $Q$ is called a {\em BRST operator\/} and the
complex $(\calh,Q)$ is called the (absolute) {\em BRST complex\/}.

As it was shown in~\cite[4.3]{KSV}, the string background defines a
sequence of $\mbox{Hom}(\calh^{\otimes n},\calh)$-valued differential
forms $\Omega_{n+1}$ on $\calp(n)$ having the
property that the integration
map $\int : CS(\calp(n))\to \mbox{Hom}(\calh^{\otimes n},\calh)$,
$\sigma \mapsto \int_\sigma \Omega_{n+1}$, makes the absolute BRST
complex $(\calh,Q)$ into an algebra over the operad
$(CS(\calp),d_S)$ of singular chains on $\calp$.

Let us recall some more or less classical definitions. Let $D$ be the
unit disk in the complex plane and let $\calf(n)$ be, for $n\geq 1$,
the space of all maps $f$ from the disjoint union of $n$ disks such
that $f$, when restricted to each disk, is the composition of the
translation and the multiplication by a nonzero complex number, and
the images of $f$ are disjoint. There is an obvious operad structure
on the collection $\calf =\{\calf(n);\ n\geq 1\}$ and the resulting
operad is called the {\em framed little disks operad\/}. The operad
$\calf$
has a suboperad $\cald$ consisting of those maps $f$ which, when
restricted to each disk, are the compositions of translations and
multiplications by positive {\em real\/} numbers. The operad $\cald$
is called the {\em little disks operad\/}. It is immediate to see
that the homotopy equivalence $\cald(n)\cong \calc_2(n)$,
where $\calc_2$ is the little 2-cubes operad discussed in
Example~\ref{positron},
given by the collapsing of disks (resp.~cubes) into a
point, induces an isomorphisms of homology operads. Therefore
$\calh(CS(\cald),d_S)\cong \calp(0,1)$. We see that the algebras
over the operad $\calh(CS(\cald),d_S)$ are exactly the Gerstenhaber
algebras and all the tricks of Example~\ref{positron} are available,
namely
every action of the operad $(CS(\cald),d_S)$ on a chain complex
induces an associate homotopy $\ssusp\mbox{\it Lie}$ structure. Let
us remark, for the sake of completeness, that the algebras over the
homology operad $\calh(CS(\calf),d_S)$ of the framed little disks
operad are so-called Batalin-Vilkovisky algebras, see~\cite{G1}.

As it was observed
in~\cite[4.7]{KSV}, there exists a natural operad map j:
$\calf \stackrel{j}{\to}\calp$.
This map is even a homotopy equivalence, but we
will not use this fact. The composition
$\cald \stackrel{i}{\hookrightarrow} \calf \stackrel{j}{\to}\calp$
combined with the string background data then induces an action of the
operad $(CS(\cald),d_S)$
on the absolute BRST complex $(\calh,Q)$. Summing up the above
remarks, we obtain the following proposition.

\begin{proposition}
String background induces a natural homotopy $\ssusp\mbox{\it Lie}$
action on the absolute BRST complex $(\calh,Q)$.
\end{proposition}
}\end{example}

\begin{example}{\rm\
\label{doutnicek}
Let $\uN(n)$ denote, for $n\geq 1$, the moduli space of stable
$(n+1)$-punctured complex curves of genus zero decorated with
relative phase parameters at double points and phase parameters at
punctures. The collection $\uN := \{\uN(n);\ n\geq 1\}$ has a natural
operad structure, see~\cite[3.4]{KSV}. A {\em closed string field
theory\/} (CSFT) consists of a string background as in the previous
example and of a choice of a smooth operad map $s :\uN \to \calp$,
see~\cite[4.4]{KSV}. In the previous example we mentioned
$\mbox{\rm Hom}(\calh^{\otimes n},\calh)$-valued forms $\Omega_{n+1}$
on $\calp(n)$; let us denote by $\overline \Omega_{n+1}$ their
restrictions $s^*(\Omega_{n+1})$ to $\uN(n)$.

Let $\calm(n)$ be the moduli space of $(n+1)$-punctured complex
projective lines~\cite[3.2]{KSV} and let $\uM(n)$ be the
Fulton-MacPherson real compactification of $\calm(n)$. We stress that
neither $\calm := \{\calm(n);\ n\geq 1\}$ nor $\uM := \{\uM(n);\
n\geq 1\}$ have an operad structure, but there still exists an operad
structure on the suspension $(\susp CS(\uM),
\susp \hskip-1mm d_S \hskip-2mm \desusp)$ of
the singular chain complex collection~\cite[3.3]{KSV}.

The BRST complex $(\calh,Q)$ of the previous example has a naturally
defined smaller subcomplex
$(\calh_{\mbox{\scriptsize \rm rel}},Q)$ called the
{\em relative BRST complex\/}. The inclusion
$\calh_{\mbox{\scriptsize \rm rel}}
\hookrightarrow \calh$ naturally splits, therefore the forms
$\overline \Omega_{n+1}$ induce
$\mbox{\scriptsize \rm Hom}(\calh_{\mbox{\scriptsize \rm
rel}}^{\otimes n},
\calh_{\mbox{\scriptsize \rm rel}})$-valued
forms $\Omega'_{n+1}$ on $\uN(n)$,
$n\geq 1$, see~\cite[4.5]{KSV}. There exists a natural projection $p:
\uN(n)\to \uM(n)$ with the fiber $(S^1)^{n+1}$ and there are
$\mbox{\rm Hom}(\calh_{\mbox{\scriptsize \rm rel}}^{\otimes n},
\calh_{\mbox{\scriptsize \rm rel}})$-valued forms
$\omega_{n+1}$ on $\uM(n)$ such
that $\Omega'_{n+1}= p^*(\omega_{n+1})$ \cite[Proposition~4.4]{KSV}.
The integration map $\int :CS(\uM(n))\to
\mbox{\rm Hom}(\calh_{\mbox{\scriptsize \rm rel}}^{\otimes n},
\calh_{\mbox{\scriptsize \rm rel}})$,
$\sigma \mapsto \int_\sigma \omega_{n+1}$,
defines on the relative BRST complex a structure of an algebra over
the operad $(\susp CS(\uM),\susp \hskip-1mm d_S \hskip-2mm \desusp)$
see~\cite[Theorem~4.5]{KSV}.

Let us recall that algebras over the homology operad
$\calh(\susp CS(\uM),\susp \hskip-1mm d_S \hskip-2mm \desusp)=
\susp \calh(\uM)$ are called
{\em gravity algebras\/}. Let us inspect more closely the
homology of the space $\uM(n)$. This space has the same homotopy type
as $\calm(n)$ and it is not difficult to see that there exists a
natural free action of the circle $S^1$ on the space $\cald(n)$ such
that $\calm(n)$ is homotopically equivalent to $\cald(n)/S^1$.
We conclude that $\susp H(\uM (n))=
\susp H(\cald(n)/S^1)$.

On the other hand, the Serre spectral sequence of the principal
fibration $S^1 \hookrightarrow \cald(n)\to \cald(n)/S^1$ gives rise
to a degree $+1$ operator $\Delta :H(\cald(n))\to H(\cald(n))$ such
that $\susp H(\cald(n)/S^1) = \Ker(\Delta)$, compare~\cite{G2}. As we
already know, the homology operad of $\cald$ is the operad for
Gerstenhaber algebras, thus $H_i(\cald(n))= 0$ for $i> n-1$ while
$H_{n-1}(\cald(n))= \ssusp \mbox{\it Lie}(n)$. As $\Delta$ has degree
$+1$, $H_{n-1}(\cald(n))\subset \Ker(\Delta)$ and we conclude that
\begin{equation}
\label{uaaaa}
\susp H(\uM(n))_i = 0
\mbox{ for $i> n-1$ while }
\susp H(\uM(n))_{n-1} = \ssusp \mbox{\it Lie}(n).
\end{equation}
Moreover, the second equation of~(\ref{uaaaa})
is compatible with the operad structures and
Proposition~\ref{trick} gives the following statement.

\begin{proposition}
\label{prstynek}
Every action of the operad
$(\susp CS(\uM),\susp \hskip-1mm d_S \hskip-2mm \desusp)$
induces an associated $\ssusp \mbox{\it Lie}$ homotopy structure.
Especially, there exists an associated homotopy
$\ssusp \mbox{\it Lie}$-structure on the relative BRST complex.
\end{proposition}

The structure predicted by the second part of the proposition was
constructed in a very explicit manner in~\cite[Corollary~4.7]{KSV}.
}\end{example}

\noindent
{\bf Final remark.}
Let us indulge at the very end of our paper in some far-stretched
remarks. We may say that two differential operads $(\cals,d_\cals)$
and $(\calt,d_\calt)$ are {\em homotopically\/} or
{\em weakly equivalent\/} if they
have isomorphic minimal models. We claim that this notion could be a
very useful tool for the study of some operads appearing in string
theory and, moreover, that we are in fact almost forced to introduce
this notion because we feel that when physicists speak about an
operad, they in fact mean an equivalence class in the above sense.
For example, we may read in~\cite[par.~3.3]{KSV} the following
sentence ($\underline{\calm}_{m+1}$ denotes the real compactification
of the moduli spaces of $n$-punctured complex projective lines -- we
already discussed this space in Example~\ref{doutnicek} where it was
denoted $\underline {\cal M}(n)$ -- and
``for instance'' was italicized by ourselves):

``Item~(5) prevents $\{\underline{\calm}_{m+1}\ |\ n\geq 2\}$ from
being an operad. But some operad structure is naturally defined on the
(singular, {\em for instance\/}) chain complexes
$C_\bullet(\underline{\calm}_{m+1})$ of $\underline{\calm}_{m+1}$'s
with coefficients in the ground field.''

This suggests that physicists would like to have the
chain functor from the category of topological
(or geometrical) operads to the
category of differential algebraic operads independent on the concrete
chains used. The only consistent way to formalize this is to
identify two operads $(\cals,d_\cals)$ and $(\calt,d_\calt)$ if there
exists a differential operad homomorphism $\phi :(\cals,d_\cals)\to
(\calt,d_\calt)$ inducing the isomorphism of homology. This is,
however, not an equivalence relation, but we may take the smallest
equivalence generated by this relation, and it can be shown that this
equivalence coincides with the homotopy equivalence relation
introduced above. As in rational homotopy theory, (minimal)
models naturally appear when we begin to study operads modulo this
relation.


\catcode`\@=11
\noindent
M.~M.: Mathematical Institute of the Academy, \v Zitn\'a 25, 115 67
Praha 1, Czech Republic,\hfill\break\noindent
\hphantom{M.~M.:} email: {\bf markl@earn.cvut.cz}

\end{document}